\documentclass[floats,floatfix,amssymb,prd,superscriptaddress,nofootinbib,preprintnumbers,notitlepage]{revtex4-1}

\usepackage{subcaption}
\usepackage{ragged2e}
\DeclareCaptionJustification{justified}{\justifying}
\makeatletter
\newcommand{\subsetsim}{\mathrel{\mathpalette\subset@sim\relax}}
\newcommand{\subset@sim}[2]{%
  \vtop{\offinterlineskip\m@th
    \ialign{\hfil##\cr
      $#1\subset$\cr\noalign{\kern0.5pt}\scalebox{0.9}{$#1\sim$}\cr
    }%
  }%
}
\makeatother

\usepackage{amssymb,amsmath,verbatim,mathtools,needspace,enumitem,etoolbox,graphicx,physics,microtype,afterpage,bm}
\usepackage[dvipsnames, usenames]{xcolor}
\definecolor{linkcolor}{rgb}{0.0,0.3,0.5}
\usepackage{booktabs}
\usepackage[unicode, colorlinks=true, linkcolor=linkcolor, citecolor=linkcolor, filecolor=linkcolor,urlcolor=linkcolor, pdfusetitle]{hyperref}
\usepackage[all]{hypcap}
\usepackage[T1]{fontenc}
\usepackage[utf8]{inputenc}
\usepackage{tabularx}
\usepackage{float}
\usepackage{cleveref}

\usepackage{multirow}
\usepackage{pifont}
\usepackage{lmodern}
\usepackage{ulem}

\allowdisplaybreaks
\usepackage{tikz}
\usepackage{framed}
\usepackage{hyperref}
\hypersetup{colorlinks, citecolor=bluscuro, linkcolor=black, urlcolor=bluscuro}
\definecolor{bluscuro}{rgb}{0.15, 0.2, .85}
\definecolor{ForestGreen}{rgb}{0.13, 0.55, 0.13}

\newcommand{\beq}{\begin{equation}}
\newcommand{\eeq}{\end{equation}}
\newcommand{\barr}{\begin{eqnarray}}
\newcommand{\earr}{\end{eqnarray}}

\newcommand{\hp}{\hat{p}}

\newcommand{\be}{\begin{equation}}
\newcommand{\ee}{\end{equation}}

\newcommand{\btheta}{{\theta}}

\newcommand{\cern}{
CERN, Theoretical Physics Department,
Esplanade des Particules 1, Geneva 1211, Switzerland}

\begin{document}

\title{
Pulsar timing array sensitivity to anisotropies in the gravitational wave background
}

\author{Paul Frederik Depta}
\email{frederik.depta@mpi-hd.mpg.de}
\affiliation{Max-Planck-Institut f\"ur Kernphysik, Saupfercheckweg 1, 69117 Heidelberg, Germany} 

\author{Valerie Domcke}
\email{valerie.domcke@cern.ch}
\affiliation{\cern}

\author{Gabriele Franciolini}
\email{gabriele.franciolini@cern.ch}
\affiliation{\cern} 

\author{Mauro Pieroni}
\email{mauro.pieroni@cern.ch}
\affiliation{\cern}

\begin{abstract}
Pulsar Timing Array (PTA) observations have recently gathered substantial evidence
for the existence of a gravitational wave background in the nHz frequency band. Searching for anisotropies in this signal is key to determining its origin, and in particular to distinguish possible astrophysical from cosmological sources. In this work, we assess the sensitivity of current and future pulsar timing arrays to such anisotropies using the full covariance matrix of pulsar timing delays. While current day pulsar timing arrays can only set mildly informative constraints on the dipole and quadrupole, we show that percent level accuracy for several low multipoles can be achieved in the near future. Moreover, we demonstrate that anisotropies in the gravitational wave background and the Hellings-Downs angular correlation, indicating the presence of GWs, are approximately uncorrelated, and can hence be reconstructed independently. 
These results can be reproduced with \href{https://github.com/Mauropieroni/fastPTA}{\texttt{fastPTA}}, a publicly available Python code to forecast the constraining power of PTA configurations.
\end{abstract}

\maketitle

\preprint{CERN-TH-2024-116} 

%{
%  \hypersetup{linkcolor=black}
%  \tableofcontents
%}
\hypersetup{
colorlinks = true,
%linkcolor = yellow,
linkcolor=blue,
citecolor= blue, %green!40!black,
urlcolor=green!40!black,%darkblue
}

%%%%%%%%%%%%%%%%%%%%%%%%%%%%%%%%%%%%%%%%%%%%%%%%%%%%%%
\section{Introduction}
\label{sec:introduction}
%%%%%%%%%%%%%%%%%%%%%%%%%%%%%%%%%%%%%%%%%%%%%%%%%%%%%%

Recent results from pulsar timing arrays (PTAs) around the world have established a first indication of gravitational waves in the nanohertz range~\cite{NANOGrav:2023gor,Antoniadis:2023ott,Reardon:2023gzh,Xu:2023wog}. The observed signal is consistent with a gravitational wave background (GWB), showing in particular evidence of an angular correlation following the Hellings-Downs (HD) function~\cite{Hellings:1983fr}, considered a smoking gun signal for gravitational wave (GW) detection at PTAs. The origin of this signal is less clear. While the most likely explanation remains a GWB emitted by supermassive black hole binaries (SMBHBs), a range of scenarios giving rise to a cosmological GWB sourced in the early Universe are currently also compatible with the data, see, e.g.,~\cite{Madge:2023cak,NANOGrav:2023hvm,Antoniadis:2023xlr,Figueroa:2023zhu,Ellis:2023oxs}. A key to distinguishing these two options lies in the search for anisotropies in the signal: while most cosmological backgrounds are expected to be highly isotropic, the GWB from SMBHBs is expected to feature anisotropies at the level of $1 - 20\%$ at low multipoles~\cite{Mingarelli:2013dsa,Taylor:2013esa,Mingarelli:2017fbe,NANOGrav:2023tcn,Sah:2024oyg,Sah:2024etc}. This raises the question of how well upcoming PTA data sets are expected to measure this quantity.

In this paper, we use the Fisher Information Matrix (FIM) formalism to forecast the sensitivity of current and future PTA experiments to anisotropies.
Our theoretical framework largely follows Ref.~\cite{NANOGrav:2023tcn} (see also~\cite{Allen:1996gp} for pioneering work on the detection of SGWB anisotropies, ~\cite{Anholm:2008wy,Mingarelli:2013dsa,Taylor:2013esa,Taylor:2015udp,Taylor:2020zpk,Ali-Haimoud:2020ozu,Ali-Haimoud:2020iyz,Pol:2022sjn} for earlier work on detecting anisotropies with PTAs, and~\cite{Gair:2014rwa,Hotinli:2019tpc} for alternative approaches), based on expanding the GW power spectrum in spherical harmonics and combining this with the geometric properties of the PTAs to obtain predictions for the time delays induced by an anisotropic GWB. For simplicity, we assume that the frequency and angular dependence of the GWB factorizes. More specialized searches can be performed assuming specific models for the GWB anisotropies, which we leave for future work. See, e.g.,~\cite{Becsy:2022pnr,Sato-Polito:2023spo,Lemke:2024cdu} for a recent dedicated search of anisotropies induced by SMBHBs and~\cite{Gersbach:2024hcc} for a more agnostic approach.
We find that the current datasets are expected to be weakly informative about anisotropies and at best able to constrain a dipole to a level of 10$\%$ while providing no non-trivial (i.e., not prior driven) constraints on the higher multipoles. These results are largely consistent with the constraints on anisotropies found in a dedicated search performed by the NANOGrav collaboration~\cite{NANOGrav:2023tcn} (NG15). We point out the impact of prior choices on this analysis, in particular for higher multipoles. Moreover, we note a slight discrepancy in the expected dipole sensitivity with respect to the constraint derived from the data and discuss possible explanations.

We demonstrate that this sensitivity is expected to increase rapidly in the future, with data including more pulsars and lower noise levels. In particular, we demonstrate that 140 pulsars (corresponding to roughly twice the number of pulsars of NG15 data release) with EPTA-like noise can set informative constraints up to multipole $\ell = 6$. Moreover, we show that, assuming SKA-like noise \cite{Janssen:2014dka,Weltman:2018zrl},  the same amount of pulsars can reach a sub-percent dipole sensitivity and can constrain $\ell = 6$ to a level of about 10\%. Reaching percent-level accuracy up to $\ell = 6$ requires about 500 pulsars with SKA-like noise.

As an aside, we clarify the role of cosmic variance for anisotropy measurements. We distinguish two effects. Firstly, the anisotropies measured in the local Universe may not represent the central values of the underlying fundamental model. As opposed to GW instruments operating at higher frequencies, this effect can be significant for PTAs since observation time and instrument size, compared to the GW frequencies measured, do not allow for many independent measurements of the GWB. Secondly, since the HD correlation plays a key role in determining the sensitivity to anisotropies, variations of the HD curve in the local universe (subject to cosmic variance) mildly impact the sensitivity to anisotropies. 
In this context, it is interesting to note that the usual expansion of the HD correlation into Legendre polynomials and the expansion of the GWB power into spherical harmonics are to good approximation orthogonal. This enables a robust reconstruction of the HD correlation in the presence of an anisotropic GWB, as well as robust measurements of the anisotropy despite the expected variance of the HD correlation. 

The remainder of this paper is organized as follows. Section~\ref{sec:anisotropies} introduces the theoretical framework describing the impact of anisotropies in the GWB on all the auto- and cross-correlations of pulsar timing delays. Based on this, Sec.~\ref{sec:fisher} introduces the FIM formalism, which can be used to estimate and forecast PTA sensitivities to anisotropic GW signals. We derive the key properties of our main results semi-analytically in the strong signal limit in Sec.~\ref{sec:strongsignal} before presenting our main numerical findings as well as their interpretation in Sec.~\ref{sec:results}. We conclude in Sec.~\ref{sec:conclusions}. Three appendices are dedicated to the derivation of the covariance matrix, the explicit expressions and properties of the generalized overlap reduction functions, and further comparisons with the NG15 results.

%%%%%%%%%%%%%%%%%%%%%%%%%%%%%%%%%%%%%%%%%%%%%%%
\section{Gravitational Wave Anisotropies in Pulsar Timing Arrays}
\label{sec:anisotropies}
%%%%%%%%%%%%%%%%%%%%%%%%%%%%%%%%%%%%%%%%%%%%%%%

\subsection{Stochastic Gravitational Wave Background}
% GWB
The spin-2 perturbations of the metric can be expressed in terms of plane GWs with frequency $f$, polarizations $\{+,\times\}$, 
and propagation directions $\hat k$ as
\begin{align}
h_{ab}(t, \vec x) =
\int {\rm d}^2\Omega_{\hat k}\>
\int_{-\infty}^{\infty}{\rm d}f\>
\left[\tilde h_+(f,\hat k) e^+_{ab}(\hat{k})
\right.
\left.
+\tilde h_\times(f,\hat k) e^\times_{ab}(\hat{k})\right]
e^{i2 \pi f(t - \hat k \cdot \vec x)}\,,
\label{e:hab(t,x)_stoch}
\end{align}
where we introduced the polarization tensors $e^{+,\times}_{ab}(\hat k)$, with the two polarizations denoted $P=\{+,\times\}$.
For a stationary and unpolarized stochastic gravitational wave background (GWB), using the convention of Ref.~\cite{Romano:2016dpx},
\begin{equation}
\langle \tilde h_P(f,\hat k) \tilde h_{P'}^*(f',\hat k') \rangle
= \frac{1}{4} {\cal P} (f,\hat k) \delta(f-f')\delta_{PP'}\delta^2(\hat k,\hat k')\,, \label{eq:hab_corr}
\end{equation}
where the factor $1/4$ arises from the definition of the one-sided power spectral density (PSD) ${\cal P}(f, \hat k)$ and the average over polarizations and the angular brackets denote the ensemble average.
We will further assume the PSD to be factorizable in frequency and angular dependent factors,  $ {\cal P} (f,\hat k) \equiv S_h(f) P(\hat k)$,
with normalization condition $\int {\rm d}^2 \Omega_{\hat k} P(\hat k) = 1 $.
Dropping the angular dependence and setting $P(\hat k) = 1/(4 \pi)$, we reproduce the well-known result for an isotropic GWB,
\begin{equation}\label{eq:omegaGWdefS}
    \Omega_{\rm GW} h^2 = \frac{h^2}{\rho_c}\frac{ {\rm d} \rho_{\rm GW}}{{\rm d} \log f}
    \equiv 
    \frac{2 \pi^2 f^3}{3 H_0^2/h^2} S_h,
\end{equation}
where $\rho_c /h^2 =  3 (H_0/h)^2 /( 8 \pi G)$ is the critical energy density of the Universe and $H_0 /h = 1/(9.78 \,{\rm Gyr})$ is the Hubble parameter today. 

To account for an anisotropic GWB, we retain the angular dependence in $P(\hat k)$ and expand in spherical harmonics as
\begin{equation}\label{eq:PkSHexpansion}
   P( \hat k )
    = 
    \sum_{\ell =0}^{\ell_{\rm max}}
    \sum_{m =-\ell}^{\ell} c_{\ell m} Y_{\ell m} (\theta, \phi),
\end{equation}
where $Y_{\ell m}(\theta, \phi)$ are the real-valued spherical harmonics, obtained from the complex-valued $\mathcal{Y}_{\ell m}$ via
\begin{equation}
Y_{\ell m}(\theta, \phi) = 
\begin{cases}
  \sqrt{2}   (-1)^m  
  \Re\{\mathcal{Y}_{\ell m}(\theta, \phi)\} & \text{if } m > 0 \\
    \mathcal{Y}_{\ell 0}(\theta, \phi) & \text{if } m = 0 \\
\sqrt{2}    (-1)^m  \Im\{\mathcal{Y}_{\ell |m|}(\theta, \phi)\} & \text{if } m < 0,
\end{cases}
\label{eq:realYlm}
\end{equation}
and normalized such that $\int {\rm d}^2 \Omega_{\hat k} Y_{\ell m}Y_{\ell' m'} = \delta_{\ell \ell'}\delta_{m m'}$. 
Notice that the normalization condition $\int {\rm d}^2 \Omega_{\hat k} P(\hat k) = 1 $ reads
\begin{equation}
 \sum_{\ell =0}^{\ell_{\rm max}}
    \sum_{m =-\ell}^{\ell} c_{\ell m} 
    \int {\rm d}^2 \Omega_{\hat k} Y_{\ell m}
    = 1. \label{eq:norm_P}
\end{equation}
Since $Y_{00} = 1/\sqrt{4 \pi}$ and  $\int {\rm d}^2 \Omega_{\hat k} Y_{\ell m} = 0$ for $\ell \geq 1$, this imposes $c_{00} = 1/\sqrt{4 \pi}$. The PSD of an isotropic GWB is thus given directly by $S_h(f)$.

\subsection{Pulsar Timing Array Response}
\label{sec:Response_ORF}
PTAs measure the timing residual $\Delta t_I$ induced by GWs along the line-of-sight between a pulsar $I$ (located in direction $\hat p_I$ at distance $D_I$) and Earth (located here at position $\vec 0$). The time shift induced on a photon reaching Earth at time $t$ then reads
\begin{equation}
    \Delta t _I = 
    \frac{1}{2} 
    \frac{\hat{p}_I^a \hat{p}_I^b}{(1+\hat{p}_I\cdot \hat k)}
    \int_0^{D_I} {\rm d} s
    \, 
h_{ab} \left( t(s),\vec{x}(s) \right) 
     \label{eq:tim_res_gen} \,, 
\end{equation}
with $s$ denoting the affine parameter parameterising the geodesic connecting and $t(s) = t - (D_I -s)$, $\vec{x} = (D_I -s) \hat{p}_I$. In practice, when extracting the timing residuals by fitting a pulsar timing model we need to additionally include a transmission function \cite{Hazboun:2019vhv}, see discussion around Eq.~\eqref{eq:t_f}. Expanding GWs as in Eq.~\eqref{e:hab(t,x)_stoch} yields
\begin{align}
\Delta t_I (t) = \int_{-\infty}^\infty {\rm d} f \> 
\int {\rm d}^2\Omega_{\hat k}\>
\sum_P
R_I^P(f,\hat k)\, \tilde h_P(f;\hat k)
\, e^{2 \pi i f t}  \,, \label{eq:tim_res_integrated}
\end{align}
where we have defined 
\begin{align}
    R_I^{P}(f,\hat k)\equiv
\frac{F^P_{\hat p_I}(\hat k)  }{i 2 \pi f}
\left(1-e^{-i 2\pi f D_I (1+\hat k\cdot\hat p_I)}\right)
\qquad 
\text{with} 
\qquad  
F^P_{\hat p_I}(\hat k) \equiv \frac12 \frac{\hp^a_I \hp_I^b}{1 + \hp_I \cdot {\hat k}} e^P_{ab}({\hat k})\,.
\label{eq:response}
\end{align}
Here we have introduced the response function $R_I^{P}(f,\hat k)$, capturing the impact of a GW with frequency $f$, polarization $P$ and propagating in the $\hat k$ direction on the signal from the $I$th pulsar. The correlation function of timing residuals from pulsars $I$ and $J$ observed at times $t_i$ and $t_j$ is then 
\begin{align}
C_{t, h, IJ, ij} &= \langle \Delta t_I (t_i) \Delta t_J (t_j) \rangle = \int_{-\infty}^\infty {\rm d}f \> \Gamma_{IJ} (f) \frac{S_h(f)}{24 \pi^2 f^2} e^{2 \pi i f ( t_i - t_j ) } \,, \label{eq:C_h_time}
\end{align}
where we have substituted Eq.~\eqref{eq:hab_corr} and we have introduced the overlap reduction function (ORF) $\Gamma_{IJ}(f)$ as 
\begin{equation}\label{eq:ORF}
\Gamma_{IJ}(f)
\equiv
\int
{\rm d}^2 \Omega_{\hat k} \, 
\kappa_{IJ}(f,\hat{k})  \,  \gamma_{IJ}(\hat k)   \, 
P(\hat k) \, ,
\end{equation}
with
\begin{equation}
\kappa_{IJ} \equiv 
\left( 1 - e^{-2 \pi i f D_I (1  + \hat p_I \cdot \hat k)}\right) \left( 1 - e^{2 \pi i f D_J (1  + \hat p_j \cdot \hat k)}  \right) \,,
\label{eq:kappa}
\end{equation}
and, following Ref.~\cite{Ali-Haimoud:2020ozu}, we have introduced the pairwise timing response function $\gamma_{IJ}(\hat k)$ as
\begin{equation}
    \label{eq:gamma}
\gamma_{IJ}(\hat k) \equiv \frac{3}{2} \sum_{P} F^P_{\hat p_I}(\hat k) F^P_{\hat p_J}(\hat k) = 
\frac{3}{4} 
\frac{\left[ \hat p_I \cdot \hat p_j 
- (\hat p_I \cdot {\hat k}) 
(\hat p_J \cdot {\hat k})\right]^2}
{(1+ \hat p_I \cdot {\hat k})(1 + \hat p_J \cdot \hat k)}
-\frac{3}{8} 
(1 - \hat p_I\cdot {\hat k})(1 - \hat p_J\cdot \hat k). 
\end{equation}
To have the overlap reduction function equal to unity if the two pulsars are identical, we have collected an overall factor $3/2$~\cite{Romano:2023zhb} in the definition of $\gamma_{IJ}$.\footnote{Notice also a factor $4$ difference compared to the notation used in~\cite{Ali-Haimoud:2020ozu} for the pairwise response function.
}
In the discussion above, we have implicitly assumed the timing model to be unbiased, i.e.\ calibrated to Minkowski space time. In practice, this may be challenging to accomplish at low frequencies. We demonstrate in App.~\ref{app:cov_dft} that this does not significantly impact the analysis in the PTA frequency band.

To simplify Eq.~\eqref{eq:C_h_time}, we notice that for PTAs, the minimum frequency is limited by the observation time $f \sim 1/T_{\rm obs} \sim  \text{nHz} \sim 0.1/\text{pc}$. Moreover, the typical distance to the pulsars is $D_I \sim~\text{kpc}$. Therefore, $f D_I \gg1$, and the rapidly oscillating pieces in Eq.~\eqref{eq:kappa} are negligible when integrated over the directions $\hat k$ (see also \cite{Mingarelli:2013dsa}). As a consequence, we can drop the terms proportional to the rapidly oscillating `pulsar terms' in $\kappa_{IJ} $. Under these assumptions we have,
\begin{align}
    \kappa_{IJ} (f, {\hat k} )\simeq 1 + \delta_{IJ} \, , \qquad \text{and} \qquad \Gamma_{IJ}(f)
    \simeq \kappa_{IJ}
    \int
    {\rm d}^2 \Omega_{\hat k} \,  \gamma_{IJ}(\hat k)   \, 
    P(\hat k) \, .
\end{align}
Here, the Kronecker delta accounts for the factor of two if the two pulsars are identical, i.e., if they have the same location in the sky and are at the same distance~\cite{Mingarelli:2014xfa}.
In this way, the ORF becomes frequency-independent. We note that for isotropic GWBs the ORF reduces to $\chi_{IJ}$, the Hellings-Downs (HD) correlation~\cite{Hellings:1983fr}, which is only a function of the angle  $\cos \zeta_{IJ} = \hat{p}_I \hat{p}_J$ between pulsars $I$ and $J$ as dictated by symmetries (see e.g. \cite{Kehagias:2024plp}) 
\begin{align}
\label{eq:HD_correlation}
\chi_{IJ} = (1+\delta_{IJ}) \left[ \frac{1}{2} + \frac{3 (1 - \cos \zeta_{IJ})}{4} \left( \ln \frac{1 - \cos \zeta_{IJ}}{2} - \frac{1}{6} \right) \right]\,.
\end{align}
On the other hand, in the more general case, anisotropies in the GW background, encoded in $P(\hat k)$, are convolved with the angular response function $\gamma_{IJ}$ in Eq.~\eqref{eq:gamma} and imprinted onto the correlation pattern of the time delays.
Expanding also the instrument response in terms of spherical harmonics with coefficients $\Gamma_{IJ,\ell m}$, 
\begin{equation}
\kappa_{IJ} \, \gamma_{IJ}(\hat k) = \sum_{\ell = 0}^{\ell_\text{max}} \sum_{m = - \ell}^\ell \Gamma_{IJ,\ell m} Y_{\ell m}(\hat k) \,,
\end{equation}
and using the orthogonality relations of the spherical harmonics, we can express the ORF as
\begin{align}
\label{eq:gamma_IJ_lm}
    \Gamma_{IJ} 
    =
    \sum_{\ell =0}^{\ell_{\rm max}}
\sum_{m =-\ell}^{\ell} c_{\ell m}  
\Gamma_{IJ, \ell m}\,.
\end{align}
The quantities $\Gamma_{IJ,\ell m}$, referred to as generalized ORFs~\cite{Mingarelli:2013dsa}, encode the geometric structure of a given PTA experiment. They can be viewed as a generalization of the HD function to higher multipoles. For explicit expressions and properties of the generalized
 ORFs see appendix~\ref{appendix:ORF}. 
In the next section, we describe how to use these quantities to estimate the sensitivity of different PTA configurations using the FIM method.

\subsection{Covariance Matrix in the Frequency Domain}
Let us conclude this section by deriving the expression for the covariance of the timing residuals in the frequency domain. For $N$ equally spaced observations between $-T_\text{obs}/2$ and $T_\text{obs}/2$ for all pulsars we can express the timing residuals $\Delta t_I (t_i)$ via the components of their discrete Fourier transform $\widetilde{\Delta t}_{I,k}$ at frequency $f_k = k/T_\text{obs}$
\begin{align}
    \Delta t_I (t_i) = \frac{1}{T_\text{obs}} \sum_k e^{i 2 \pi f_k t_i} \widetilde{\Delta t}_{I,k}\,,
\end{align}
where $f_k$ runs over both positive and negative frequencies.
As we show in appendix~\ref{app:cov_dft}, for large $N$ the discrete Fourier components $k$ and $l$ from pulsars $I$ and $J$ have the covariance matrix\footnote{The prefactor $T_\text{obs}^2 / N^2$ is due to the scaling of the covariance matrix with the number of observations squared,
and to match the dimensions of~\cite{Babak:2024yhu}.}
\begin{align}\label{eq:Chfinal}
C_{h,IJ,kl} &= \frac{T_\text{obs}^2}{N^2} \sum_{m, n} e^{-i 2 \pi (f_k t_m - f_l t_n)} \langle \Delta t_{I} (t_m) \Delta t_J (t_n) \rangle \\
&\simeq T_\text{obs}^2 \int_{-\infty}^\infty {\rm d} f \> \Gamma_{IJ} \frac{S_h(f)}{24 \pi^2 f^2}  \, \mathrm{sinc} \left[ \pi T_\text{obs} (f - f_k) \right] \,  \mathrm{sinc} \left[ \pi T_\text{obs} (f - f_l) \right]  \,. 
\end{align}
If $S_h(f)/f^2$ is approximately constant, this integration can be performed analytically. At this point we can also switch to using positive frequencies $f_k = k/T_\text{obs}$ only and obtain,
\begin{align}
C_{h,IJ, kl} &
\simeq \Gamma_{IJ}(f_k) \frac{S_h (f_k)}{12 \pi^2 f_k^2} \, 
T_\text{obs} \, \mathrm{sinc} [\pi T_\text{obs} (f_k-f_l)]
\simeq \Gamma_{IJ}(f_k) \frac{S_h (f_k)}{12 \pi^2 f_k^2} \delta(f_{k} - f_{l})\,. \label{eq:cov_fourier_approx}
\end{align}
For a discussion of the accurateness of these approximations, in particular regarding possible correlations across different frequencies, see  appendix~\ref{app:cov_dft}.

In a PTA, the signal extraction involves computing timing residuals by subtracting the anticipated time of arrival determined by the timing model. Fitting for the parameters entering the timing model (such as the intrinsic pulsar rotation frequency, its derivative, and proper motion), results in a polynomial suppression of sensitivity towards lower frequencies.
Notably, this suppression predominantly occurs below the frequency $1/T_{\rm obs}$.\footnote{Improved sensitivity to the GWs with frequencies below $1/T_\text{\tiny obs}$ can be achieved when including
higher order spin-down terms of the pulsar timing model which lead to slow variations of $\Delta t(t)$ (see, e.g.,~\cite{DeRocco:2022irl,DeRocco:2023qae}).}
The attenuating effect on the signal can be effectively quantified by a transmission function exhibiting a $1/f^6$ scaling (for a quadratic spin-down model) below the frequency of $1/T_{\rm obs}$~\cite{Hazboun:2019vhv}. 
We model this transmission function as\footnote{An additional reduction in sensitivity occurs approximately at $f=1 /{\rm yr}$ due to the inclusion of sky position and proper motion in the fitting process, and at $f = 2/ {\rm yr}$ owing to parallax. These adjustments result in conspicuous spikes within the sensitivity profile. For example, if the pulsar resides within a binary system characterized by a period encompassing the frequency range of interest, the transmission function experiences a dip attributed to the incorporation of orbital parameters into the fitting procedure.
Given that these sensitivity reductions primarily manifest at relatively higher frequencies, they are deemed negligible for our purposes.}
    \begin{equation}
        {\cal T}(f) \simeq \left[  1+ 
        1/\left (f T_\text{\tiny obs} \right ) \right ]^{-6}.
    \label{eq:t_f}
    \end{equation}
This transmission function needs to be included in the covariance matrix in Eq.~\eqref{eq:cov_fourier_approx}. Further allowing for different observation times $T_I$ for each pulsar, leading to an effective overlapping time $T_{IJ}={\rm min}[T_{I},T_{J}]$ between pulsars $I$ and $J$, we model the covariance matrix as
\begin{align}
 C_{h,IJ, kl}
\simeq 
\left [
\frac{{\cal T}_I(f)
{\cal T}_J(f)
T_{IJ} }{T_\text{obs} }
\right]^{1/2}
\Gamma_{IJ}(f_k) \frac{S_h (f_k)}{12 \pi^2 f_k^2} \delta(f_{k} -f_{l})\,.
\label{eq:CIJ}   
\end{align}

\section{Fisher forecast of sensitivities}
\label{sec:fisher}
Let us start by expressing the full covariance matrix for the timing residuals from pulsars $I$ and $J$ in terms of the noise and GW signal contributions (which are assumed to be uncorrelated) as
\begin{equation}\label{e:C_IJ}
C_{IJ} = C_{n,IJ} + C_{h,IJ}\,,
\end{equation}
where, as discussed above, for the signal part we expand the ORF in spherical harmonics as 
\begin{align}
C_{h,IJ}(f) 
& = 
\left [
\frac{{\cal T}_I(f)
{\cal T}_J(f)
T_{IJ} }{T_\text{obs} }
\right]^{1/2}
\left( \frac{S_h(f)  }{12 \pi^2 f^2}\right)
\times 
 \sum_{\ell =0}^{\ell_{\rm max}}
\sum_{m =-\ell}^{\ell} c_{\ell m} 
 \Gamma_{IJ,\ell m}\,.
\label{eq:full_covariance}
\end{align}
This covariance matrix extends the one considered in Ref.~\cite{Babak:2024yhu} as it accounts for an arbitrary GWB angular dependence, i.e., it does not rely on the assumption of an isotropic GWB. 
To describe $C_{n,IJ}$, we adopt the noise modelling from~\cite{Babak:2024yhu}, which assumes the noise to be uncorrelated for different pulsars. 
In practice, this corresponds to approximating $C_{n,IJ} = \delta_{IJ} P_{n,I}$, where $P_{n,I}$ includes white noise from time of arrival match filtering (typically controlled by the parameters EQUAD, ECORR, and EFAC in PTA analyses~\cite{EPTA2-pulsars,NG15-pulsars}), red noise from stochasticity in pulsar rotation, and chromatic noise components from temporal variations in dispersion measure and scattering variations. 
In Ref.~\cite{Babak:2024yhu} it was shown that uncertainties on the GWB spectral parameters derived assuming this noise model correctly reproduce the ones obtained in the EPTA analysis \cite{EPTA2-pulsars}.
We will simulate future PTA datasets, by randomly sampling noise parameters from their distribution built out of current observations, and we will denote this as ``EPTA-like'' noise.

Assuming the data $\tilde{d}^k$, with $k$ running over frequencies $f_k$, to be Gaussian variables with zero means and described only by their variance $C_{IJ}(f_k,\btheta)$, the log-likelihood can be written as
\cite{Contaldi:2020rht,Bond:1998zw,Babak:2024yhu}
\begin{equation}\label{eq:likelihood}
- \ln \mathcal{L}
(\tilde{d} \vert \btheta )
= {\rm const. }+
\sum_{k,IJ} 
 \ln \left [ C_{IJ}(f_k,\btheta)\right ] 
+ \tilde{d}_I^k C_{IJ}^{-1} (f_k,\btheta) \tilde{d}_J^{k*}  \; ,
\end{equation}
where $\btheta$ denotes the set of model parameters, including the  coefficients $c_{\ell m}$ parametrizing the anisotropies in the GWB. The Fisher Information Matrix (FIM) is given by the second derivative of the log-likelihood with respect to the model parameters evaluated at the best fit for the parameters (which to perform forecasts, is assumed to match with injection). As shown, e.g., in Ref.~\cite{Bond:1998zw}, the FIM is given by
\begin{equation}
    F_{\alpha \beta} 
= 
 \sum_{f_k, IJKL}
 C^{-1}_{IJ} C^{-1}_{KL} 
\frac{\partial C_{h,JK}}{\partial \theta_\alpha}
\frac{\partial C_{h,LI}}{\partial \theta_\beta}\,.
\label{eq:fisher}
\end{equation}
Assuming Gaussian distributions, from this quantity, we can easily obtain the covariance matrix $C_{\alpha \beta}$ for the model parameters $\theta$ by inverting the FIM, i.e., $C_{\alpha \beta} = (F^{-1})_{\alpha \beta}$. 
This directly gives a measure of the expected sensitivity to the model parameters (see, e.g.,~\cite{Coe:2009xf}).
Finally, given a model, the priors on the model parameters should be imposed consistently to forecast the results of a full Bayesian analysis. While Gaussian priors can be trivially included in the FIM formalism, adding constant terms to the FIM diagonal, including informative flat priors (e.g., hard cuts) requires a more elaborate procedure. A simple strategy to enforce flat priors in the FIM framework relies on rejection sampling. For this purpose, ``measurements'' are generated using a normal distribution centered in the injection parameters and with covariance given by $C_{\alpha \beta}$, and values lying outside the prior ranges are rejected. This procedure, which provides a fast, but reasonably accurate way, to enforce flat priors in FIM forecasts, will be used in the following when such priors are considered.

\section{Semi-analytical results in the strong signal limit}
\label{sec:strongsignal}

\begin{figure*}[t]
\centering
\includegraphics[width=0.49\textwidth]{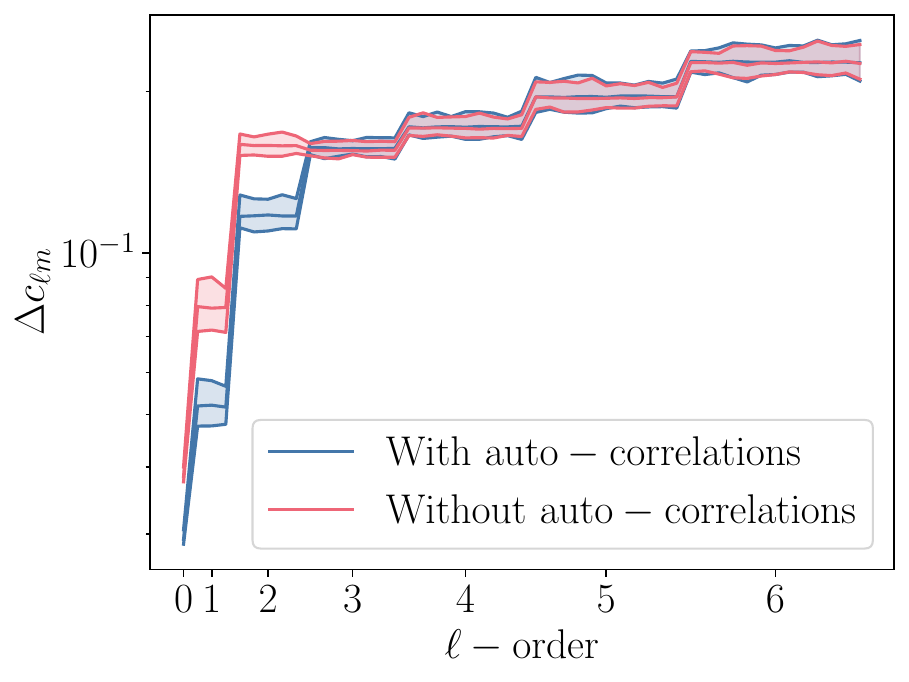}
\includegraphics[width=0.49\textwidth]{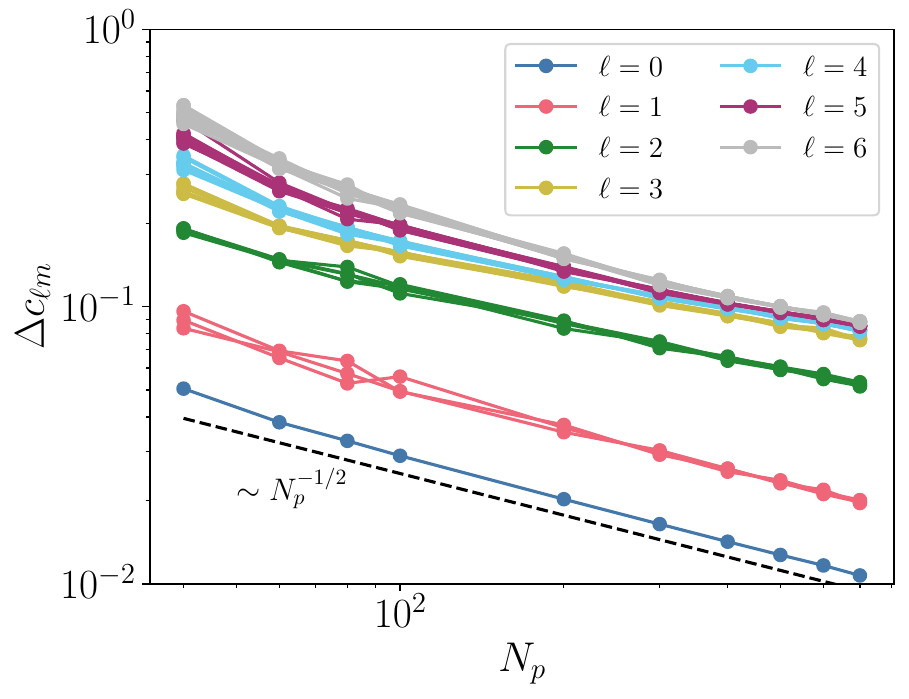}
\caption{Sensitivity to anisotropies in the strong signal limit for multipoles $\ell = 0,\dots,6$ assuming an isotropic injection. We report 1$\sigma$ uncertainties on spherical harmonics coefficients $\Delta c_{\ell m}$ assuming $N_f = 1$.
\textit{Left panel:} the colored band indicates 
95\% C.I. obtained by varying over many realizations of $N_p = 100$ pulsars sampled from an isotropic distribution on the sky. 
We indicate with different colors the case in which autocorrelations (i.e., $I=J$ terms) are retained (blue) or dropped (red). 
\textit{Right panel:} Scaling of the uncertainties with the number of pulsars $N_p$. Multiple lines for each $\ell$ correspond to different values of $m$, which overlaps in the limit of large $N_p$ due to the assumed statistical isotropy.  }
\label{fig:Deltaclm}
\end{figure*}

In order to gain an intuition of the behaviour of the FIM and thus the expected sensitivity to anisotropies, we start by neglecting the noise contribution to the covariance matrix, corresponding to the strong signal limit. Assuming equal observation time for all pulsars (i.e., equal $T_{IJ} = T_{\text{obs}}$ and ${\cal T}_I = {\cal T}$ for all $I,J$ pulsars indexes), the frequency-dependent contribution factorize out.
Let us further assume the GWB to be isotropic, i.e., $c_{00} = 1/\sqrt{4 \pi}$ and $c_{\ell m} = 0$ for all other multipoles. Then,
\begin{align}
 C_{IJ}(f) \simeq \sqrt{4 \pi} \, {\cal T} \left( \frac{S_h(f)  }{12 \pi^2 f^2}\right) 
  \Gamma_{IJ,00}
 \,, \quad 
 \text{and}
 \quad 
 \frac{\partial C_{IJ}(f)}{\partial c_{\ell m}} \simeq {\cal T}   \left( \frac{S_h(f)  }{12 \pi^2 f^2}\right)  
 \Gamma_{IJ,\ell m}\,,
\end{align}
where $\Gamma_{IJ,00}$ is the HD function depending only on the angular distance between $\hat p_I$ and $\hat p_J$. Inserting these equations into the FIM expression in Eq.~\eqref{eq:fisher} yields
\begin{equation}
F_{\ell m,\ell^\prime m^\prime} 
=
4 \pi
N_f
\Gamma^{-1}_{IJ,00}
\Gamma^{-1}_{KL,00}
\Gamma_{JK,\ell m}
\Gamma_{LI,\ell^\prime m^\prime}\,,
 \label{eq:strong-signal}
\end{equation}
where $N_f$ denotes the number of frequency bins.
For a given set of pulsars the generalized ORFs $\Gamma_{IJ, \ell m}$ can be easily computed using Eq.~\eqref{eq:gamma_IJ_lm}. Inverting the resulting FIM, we can estimate the ability to set upper bounds on the coefficients $c_{\ell m}$ parametrizing the anisotropy in this limit - assuming the observed GWB is isotropic.\footnote{We present results assuming anisotropic injections in the appendix. See, in particular, Fig.~\eqref{fig:Deltaclm_anisotropies}.} This is depicted in Fig.~\ref{fig:Deltaclm}. 
We note the remarkably good sensitivity to the dipole $\ell=1$, whereas at higher multipoles the sensitivity decreases significantly, indicating that large pulsar timing arrays with excellent noise control will be needed to set meaningful constraints.

\begin{figure*}[t]
\centering
\includegraphics[width=0.245\textwidth]{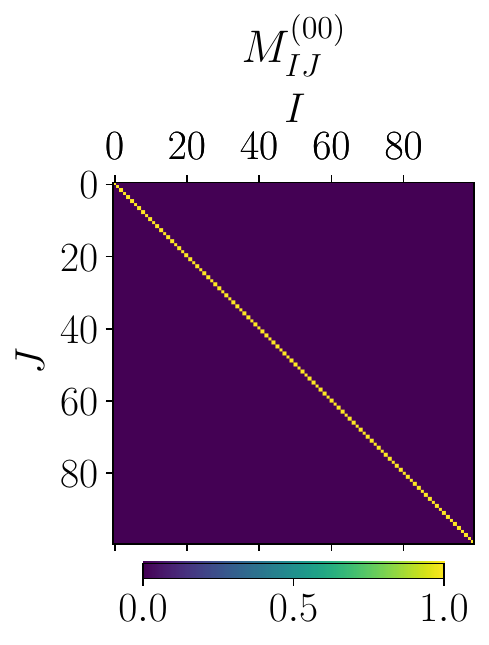}
\includegraphics[width=0.245\textwidth]{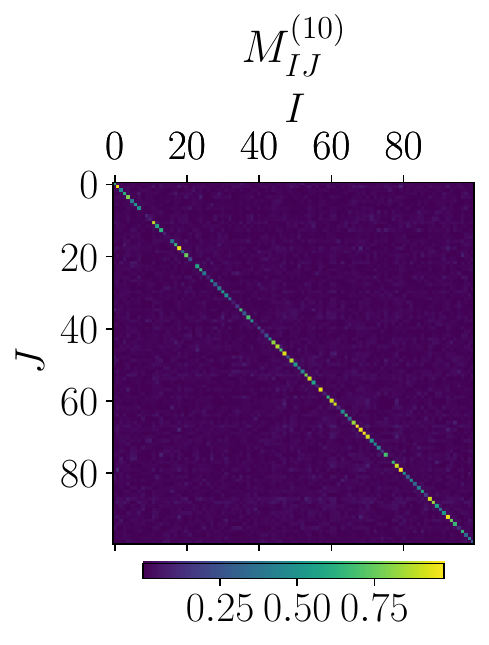}
\includegraphics[width=0.245\textwidth]{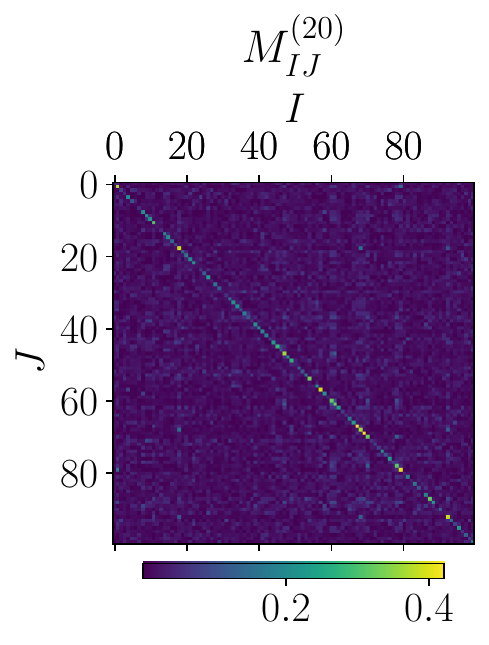}
\includegraphics[width=0.245\textwidth]{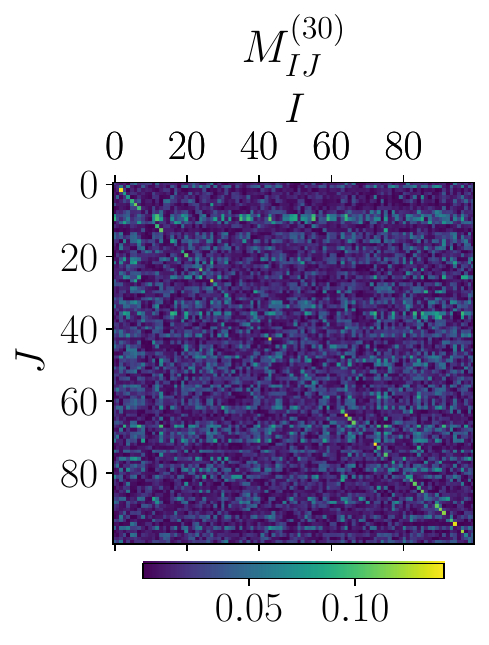}
\caption{Structure of $M_{IJ} = \Gamma_{IK,00}^{-1} \Gamma_{KJ, \ell m}$ for different multipoles $\ell = 0,..,3$, illustrating that at low multipoles ($\ell = 0,1$) this matrix is (approximately) diagonal in the pulsar indices $I,J$, whereas at larger multipoles this structure is becomes increasingly lost. This figure is generated by evaluating $M_{IJ}$ numerically for 100 randomly distributed pulsars.
We chose $m=0$ but similar results are obtained for the remaining values of $m$.}
\label{fig:MIJ}
\end{figure*}

The features revealed in the strong signal limit can be traced back to the convolution of the multipole expansion of the GWB with the angular structure of the pairwise timing response function $\gamma_{IJ}$. To illustrate this, we introduce $M^{(\ell m)}_{IK} = \Gamma_{IJ,00}^{-1} \Gamma_{JK, \ell m}$, such that the diagonal entries of Eq.~\eqref{eq:strong-signal} read\footnote{The focus on the diagonal elements is justified by the approximately diagonal structure of the FIM in this basis, see Fig.~\ref{fig:HD_anis_cov}.}
\begin{align}
    \label{eq:FIM_M_mat}
    F_{\ell m, \ell m} = 4 \pi 
    N_fM^{(\ell m)}_{IK} M^{(\ell m)}_{KI} \,.
\end{align}
For the monopole, $\ell = 0$, $M^{(00)}_{IK}$ is simply the $N_p$-dimensional unit matrix. Squaring and tracing over this matrix gives $F_{00,00} = 4 \pi N_f N_p$, which implies that $\Delta c_{00}$ should scale as $N_p^{-1/2}$, as observed in Fig.~\ref{fig:Deltaclm}. For the dipole, $\ell = 1$, we note that both $\Gamma_{IJ,00}^{-1}$ and $ \Gamma_{JK, 1 m}$ are approximately diagonal in the pulsar indices. This can be understood from the explicit expressions for these quantities. Recalling that $\Gamma_{IJ,00}$ is the HD function, we see immediately that the strongest support is at $\zeta_{IJ} = 0$, where both the auto-correlation term and cross-correlations of pulsars with small opening angle contribute constructively. A similar situation arises for the dipole ORF $\Gamma_{IJ,1 m}$ as shown explicitly in the appendix \ref{appendix:ORF}.\footnote{Fig.~\ref{fig:Gamma_real} in the appendix displays $\Gamma_{IJ,1 m}$ in the computational frame, where one of the pulsars is located on positive $z$-axis and the other on the $(\hat x, \hat z)$ plane. Rotating to allow for arbitrary angles for both pulsars mixes the $m$-components, but conserves on average the strongest support at $\zeta_{IJ} = 0$ for low multipoles.} Consequently, the matrix $M^{(1m)}_{IK}$ is still approximately diagonal with elements of similar magnitude, as illustrated in Fig.~\ref{fig:MIJ}. This implies that the dipole features a similar scaling as the monopole, $c_{1m} \propto N_p^{-1/2}$.

As we go to higher multipoles, the approximately diagonal structure of the generalized ORFs $\Gamma_{IJ,\ell m}$ is lost, resulting in a partial loss of the diagonal structure in $M^{(\ell m)}_{IK}$, see Fig.~\ref{fig:MIJ}. Taking into account that in general $M^{(\ell m)}_{IK}$ is not a symmetric matrix, this leads to partial cancellations when computing the trace in the FIM resulting in a milder scaling in the number of pulsars. We can interpret this as `self-noise' of the GWB multipoles correlated through the HD function. This is precisely what we observe for higher multipoles at large $N_p$ in the right panel of Fig.~\ref{fig:Deltaclm}. The limiting case of random entries in $M_{IK}$ would imply a $N^{-1/4}$ scaling in $\Delta c_{\ell m}$. The scaling observed in the right panel of Fig.~\ref{fig:Deltaclm} falls between these two limiting cases. 

This picture does not change significantly when considering an anisotropic GWB. An anisotropic injection enters the FIM through the $C_{IJ}^{-1}$ terms in Eq.~\eqref{eq:fisher}. However, this contribution is always at most comparable to the HD function. This explains why even in the presence of an anisotropic GWB, we expect the accuracy of the anisotropy measurement to scale as in Fig.~\ref{fig:Deltaclm} with the number of pulsars. We confirmed this scaling numerically by performing Fisher forecasts with an anisotropic GWB injection, both in the strong signal limit (shown in Fig.~\ref{fig:Deltaclm_anisotropies} in appendix \ref{app:anisotropic injection SSL}) and including the pulsar noise.

Albeit very much simplified, this exercise revealed some key properties of the expected sensitivity of PTAs to anisotropies. In particular, we note that the geometric structure of PTAs leads to an increased sensitivity to low multipoles, which moreover features a stronger scaling with the number of pulsars. This is partially, but not entirely, driven by the contributions from pulsar auto-correlations. This can be observed explicitly by looking at the left panel of Fig.~\ref{fig:Deltaclm}, where both the results including and excluding auto-correlations are shown. Auto-correlations contribute to $\Gamma_{IJ, \ell m}$ up to $\ell \leq 2$, as can be easily noticed by inspecting $\gamma_{II} \propto (1-\hat p_I \cdot \hat k)^2 $ \cite{Ali-Haimoud:2020iyz}. Therefore, their inclusion in the analysis increases the available information on the low multiple anisotropies, provided noise is not dominant in these terms. 
In the following section, we return to these questions in more detail, taking into account realistic noise models and pulsar configurations.

\section{Results}
\label{sec:results}
Following Ref.~\cite{NANOGrav:2023tcn}, we show our results obtained in the spherical harmonic
basis in terms of $C_\ell$, which is the squared angular power in each multipole mode $\ell$
\begin{equation}
    C_\ell = \frac{1}{2\ell +1 } 
    \sum_m |c_{\ell m}|^2.
\end{equation}

\subsection{Imposing priors on $C_\ell / C_0$}

The power contained in the GWB in each direction is physically constrained to be a positive quantity. An expansion in spherical harmonics with free coefficients is a priori not subject to this constraint. Consequently,
while positivity of power in any direction of the sky is automatically respected by any physical GWB, the reconstruction of the spherical harmonic expansion's coefficients may lead to unphysical results for parts of the posterior. This is particularly true if only a low number of multipoles are included in the reconstruction and if the uncertainties on the reconstructed parameters are large.

\begin{figure*}[t]
\centering
\includegraphics[width=0.49\textwidth]{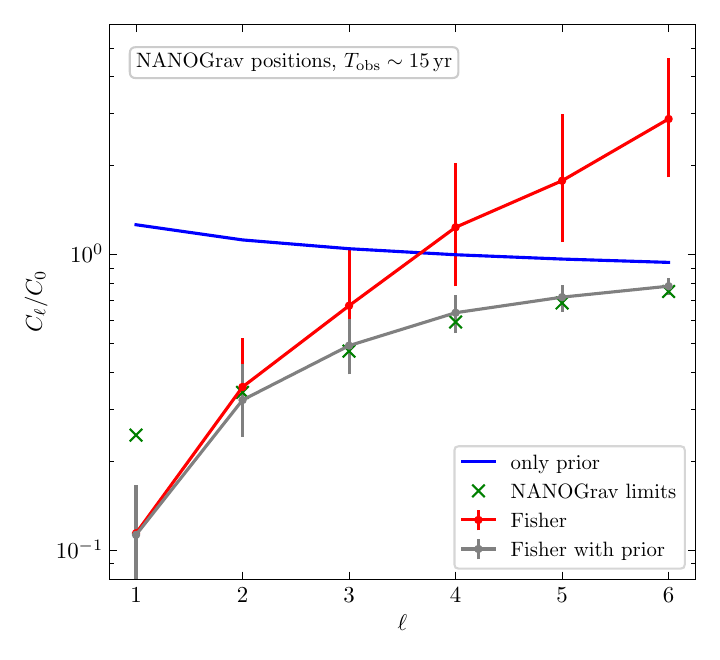}
\includegraphics[width=0.49\textwidth]{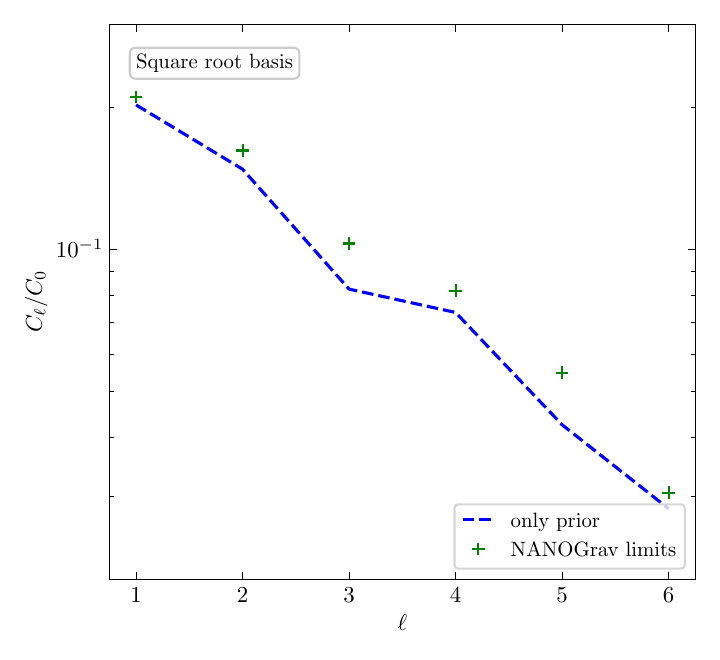}
\caption{
\textit{Left:} Mean and 95\% confidence interval for 95\% C.L. upper bounds on $C_\ell / C_0$ in linear basis from the FIM without (red) and with (grey) restriction to prior $|c_{\ell m}| \leq 5 / (4 \pi)$ as in~\cite{NANOGrav:2023tcn} for 68 pulsars at the same positions as in the NANOGrav 15 yr data set with $T_\text{obs}\sim 15 \, \mathrm{yr}$. The C.I. accounts for sampling the pulsar noise parameters from their distributions \cite{Babak:2024yhu}. In blue we show the results just drawing from this uniform prior and in green the limits from~\cite{NANOGrav:2023tcn}. \textit{Right:} 95\% C.L. upper bounds on $C_\ell / C_0$ from drawing from uniform priors $|g_{LM}| < 50$ with $g_{00} = 1$ in the square root basis (blue dashed) together with limits from~\cite{NANOGrav:2023tcn} (green).}
\label{fig:ng_comparison}
\end{figure*}

Imposing priors motivated by positivity may speed up data analysis, as unphysical regions of parameter space are not explored in the analysis. On the other hand, the leakage of posteriors into unphysical regions may simply indicate a lack of constraining power in the data, in which case the impact of priors on the posteriors needs to be checked particularly carefully.
Let us revisit the two methods implemented by the NANOGrav collaboration in Ref.~\cite{NANOGrav:2023tcn} in view of this.
\begin{itemize}
    \item {\it Spherical harmonics basis.} Following the analysis performed in Ref.~\cite{NANOGrav:2023tcn}, we restrict the maximal value of the coefficients parametrizing the anisotropy in the GWB to $|c_{\ell m}| < 5/(4 \pi)$, where the factor $4 \pi$ reflects a difference in normalization choice compared to Ref.~\cite{NANOGrav:2023tcn}. 
    These priors are included in the analysis as described at the end of the previous section. The results are depicted as the gray curve in the left panel of Fig.~\ref{fig:ng_comparison}. 
    For comparison, we show also the results obtained without imposing this prior (red), and the constraint on the parameter space obtained from only the prior (blue), together with the limits found in Ref.~\cite{NANOGrav:2023tcn} (green points).
    \item {\it Square root basis.} We perform a change of basis to the so-called square-root-basis where
\begin{equation}\label{eq:sqrtBasis}
   P( \hat k )
    = 
    \left [\sum_{L=0}^{L_{\rm max}}
    \sum_{m =-L}^{L} g_{LM} Y_{LM}
    \right ]^2.
\end{equation}
By construction, $P(\hat k)$ is now positive for any choice of $\{g_{LM}\}$. 
It is possible to relate the coefficients in the two bases, $c_{\ell m}$ and $g_{LM}$, through an infinite summation
\begin{align} \label{eq:clm_from_alm}
c_{\ell m} &= (-1)^m \sum_{L=0}^{\infty}\sum_{M=-L}^{L}\sum_{L'=0}^{\infty}\sum_{M'=-L}^{L'} g_{LM}g_{L'M'} 
\sqrt{\frac{(2l+1)(2L+1)(2L'+1)}{4\pi}} 
\begin{pmatrix} \ell & L & L' \\ -m & M & M'\end{pmatrix}\begin{pmatrix} \ell & L & L' \\ 0 & 0 & 0\end{pmatrix}\,,
\end{align}
in terms of the Clebsch-Gordon coefficients. 
As in Ref.~\cite{NANOGrav:2023tcn}, we truncate this summation at $L_\text{max} = \ell_\text{max}/2$ where $\ell_\text{max} = 6$. See also Refs.~\cite{Taylor:2020zpk,Banagiri:2021ovv} for a discussion of the accuracy of this procedure. Drawing the coefficients $g_{LM}$ from a uniform prior with $|g_{LM}| < 50$ with $g_\text{00} = 1$,\footnote{
This normalization differs from the normalization condition $c_{00} = 1/\sqrt{4\pi}$ implemented in the spherical harmonic basis. This difference drops out when considering the quantity $C_\ell/C_0$.}
we can estimate the prior-only constraint on the $c_{\ell m}$, and hence $C_\ell$, parameter space. This is shown as the dashed blue curve in the right panel of Fig.~\ref{fig:ng_comparison}. For comparison, the green points indicate the limits found in Ref.~\cite{NANOGrav:2023tcn}. As already cautioned in Ref.~\cite{NANOGrav:2023tcn},  we find these limits to reflect only the information on the prior.
Moreover, since the expansion in this basis is non-linear in the basis functions $Y_{LM}$, implementing the Fisher matrix becomes less straightforward. 
For these reasons, we do not pursue this method any further in this work.
\end{itemize}
Once the data becomes sufficiently constraining, all methods should of course lead to the same results, independent of prior choices or methodology. Nevertheless, different methods may be better adapted and more efficient to address different questions. In the remainder of this work, we will work on the spherical harmonics basis.

%%%%%%%%%%%%%%%%%%%%%%%%%%%%%%%%%
\subsection{Forecasts with current PTA configurations}

Fig.~\ref{fig:ng_comparison} illustrates the impact of prior choices on a mock data sample using 68 pulsars at the same positions as in the NANOGrav 15 year data set in the spherical harmonic bases (left) and the square root basis (right). In both cases, the green crosses indicate the limits found by the NANOGrav collaboration in~\cite{NANOGrav:2023tcn} whereas the blue lines indicate the limits found by sampling the prior-only limited parameter space. We note that for the square-root basis (right) this can to good accuracy explain the limits quoted in~\cite{NANOGrav:2023tcn}, indicating the absence of constraining power in the data compared to the adopted prior. 

On the other hand, in the spherical harmonic basis, the prior imposed in~\cite{NANOGrav:2023tcn} is less constraining and we do confirm the expectation of a data-driven limit for low $\ell$. Performing a more detailed Fisher forecast analysis without (red) and with (gray) priors for $10^3$ different isotropic GWB and noise realizations we find that the flat prior on the $c_{\ell m}$ described above significantly impacts the constraints for $\ell \geq 3$. Conversely, the constraints for $\ell = 1$ and $\ell = 2$ seem data-driven. Overall, we find good agreement between our estimated sensitivity and the limits quoted in~\cite{NANOGrav:2023tcn}, validating the procedure adapted in this work and enabling us to proceed with forecasting future sensitivities in the next section.

However, we do note that our estimated sensitivity to the dipole lies somewhat below the limit quoted in~\cite{NANOGrav:2023tcn}. In light of our considerations based on the strong signal limit above, possible explanations for this discrepancy are: 
\begin{itemize}
\item[(i)] Our noise model is based on~\cite{Babak:2024yhu}, this may underestimate the noise in the dataset of~\cite{NANOGrav:2023tcn}.
As higher multipoles are prior-dominated, an underestimation of an isotropic noise component would mainly impact the analysis of the low multipoles, which could partially explain the discrepancy. To illustrate this, we show results for an artificially enhanced white noise spectrum in the left panel of Fig.~\ref{fig:ng_comparison2} in the appendix. We see that enhancing the noise level to the point of rendering the data sensitivity compatible with observations leads to a degradation of the sensitivity at higher multipoles, which, in turn, seems in tension with the reported limits. Hence, we conclude that this can at best be a partial explanation. 
Also, we stress again that in the current analysis, we assume the noise to be diagonal in $(IJ)$ and the noise properties of each pulsar are sampled from the same distribution, which means the noise is taken to be statistically isotropic across all pulsars. As a consequence, there are no correlated noise components related to the detection on Earth. This assumption may be violated in the NANOGrav dataset due to an imperfect timing model.
This effect could mimic an anisotropic noise, explaining a reduced sensitivity to the dipolar anisotropy.
\item[(ii)] Our Fisher forecast assumes Gaussian distributions for $c_{\ell m}$ and reconstructs ensemble averages. Non-linearities in the data are not captured, which requires particular caution with statements that refer to the tails of these distributions and hence the edges of the posterior distributions. 
\item[(iii)] Moreover, in a given PTA data set, not all pulsars are equally informative. Thus, only a few high-quality pulsars may be available to test large scales and hence low multipoles.
\item[(iv)] This may be a first indication of a non-vanishing dipole component in the PTA data. As we show in the right panel of Fig.~\ref{fig:ng_comparison2} in the appendix, this explanation would require a dipole contribution with $C_1/C_0 \gtrsim 2\%$. We expect upcoming data releases of the different PTA collaborations to shed more light on these questions.
\item[(v)] Finally, we emphasize that our analysis is based on sensitivity forecasts only, whereas Ref.~\cite{NANOGrav:2023tcn} reports the result of an analysis on real data. The latter is a much more challenging problem, both due to the complexity of realistic data and the computational requirements of the Bayesian analysis. Our sensitivity estimates thus represent an idealized target.
\end{itemize}

In case the discrepancy observed in the dipole remains in future PTA analyses, one could search for evidence of the GW nature of the dipolar excess, compared to unmodelled noise systematics, by inspecting the reconstructed dipolar ORF. Similarly to what is currently done for the HD correlation (i.e. $\ell = 0$), 
one could parametrise $\Gamma_{1m}$ in the computational frame in a suitable basis (see more details in appendix \ref{appendix:ORF}) and test whether the shape of the reconstructed ORF is compatible with the GW prediction. For example, one could decompose $\Gamma_{1m}$ into Legendre polynomials, to check that the measured power comes from a GW.

\subsection{Forecasts for future PTA configurations}

Future PTA configurations will contain significantly more pulsars, and longer observation times, while simultaneously achieving lower noise levels.
In the upcoming years, the Square Kilometer Array (SKA) will be a central protagonist among PTA collaborations~\cite{Janssen:2014dka,Weltman:2018zrl}. At the moment, PTA datasets are mostly limited by radiometer noise. The SKA will provide higher-precision pulsar timing measurement with uncertainties below $\sim$100ns~\cite{Janssen:2014dka}. This would be roughly 10 times better than what is achieved by current generation telescopes \cite{Lazio:2013mea} and would allow increased sensitivity at currently observed frequencies while extending the range of informative frequencies (i.e., those that are not noise dominated) to higher frequencies.

\begin{figure*}[t]
\centering
\includegraphics[width=1\columnwidth]{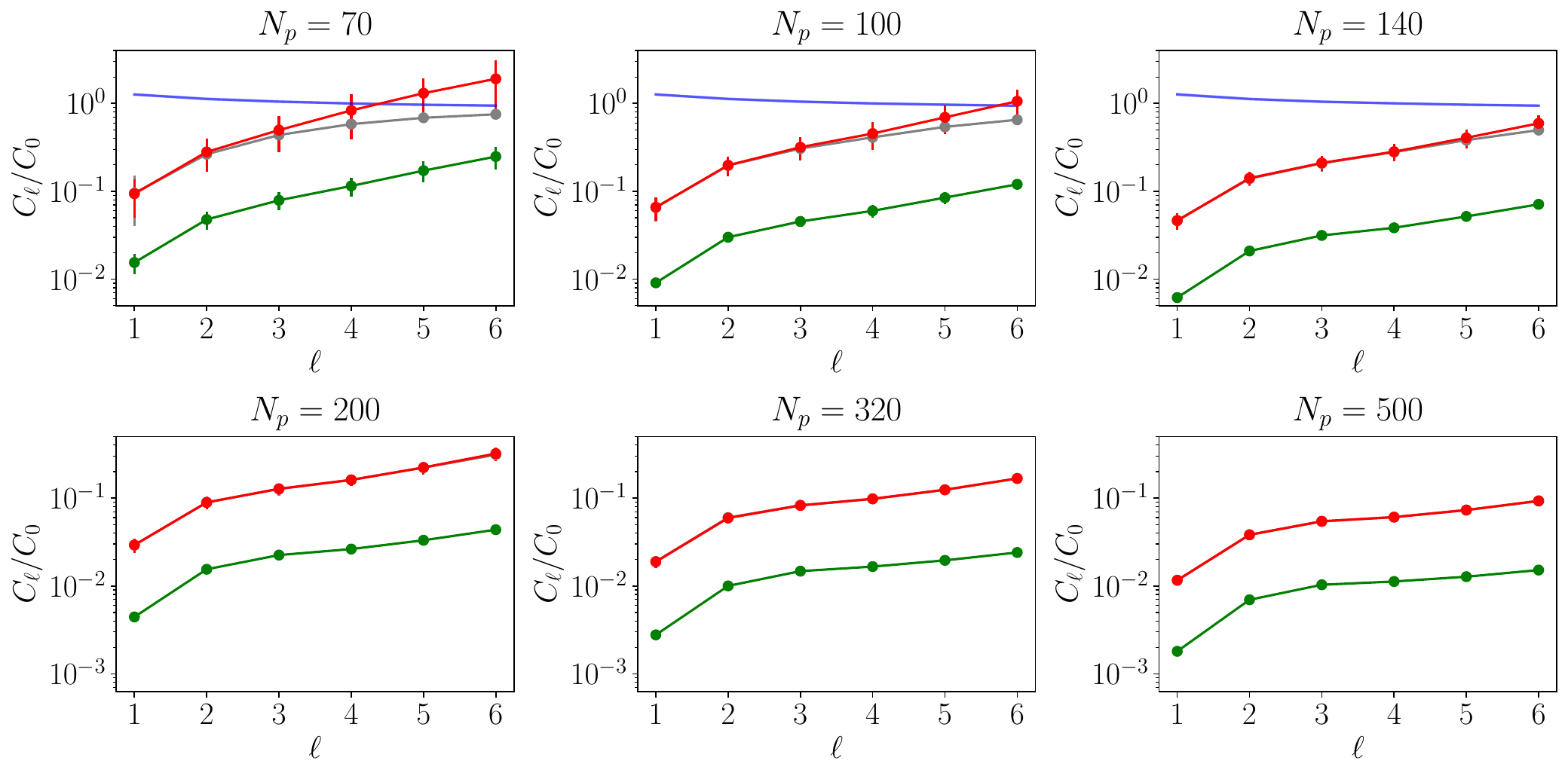}
\caption{
95 \% C.L. upper bounds on $C_\ell$ assuming EPTA-like noise (red) and SKA-like noise (green) for different numbers of pulsars and $T_{\rm obs} = 15 {\rm yr}$. We sample 10 realizations of noise and pulsar locations in the sky (with uniform distribution on the sphere). 
The gray lines indicate the result adopting the prior $|c_{\ell m}| < 5/(4 \pi)$, while the blue reports the upper bound just drawing from the same uniform prior. In the bottom row, due to the increased sensitivity, such a prior plays no role. }
\label{fig:forecast}
\end{figure*}

In this section, we use the methods developed above to estimate the sensitivity of future PTA configurations to anisotropies of the GWB. 
Starting from `current day' PTAs with 70 pulsars randomly distributed in the sky with $T_{\rm obs} = 15 {\rm yr}$, we simulate future configurations with up to 500 pulsars. The result assuming current-day (EPTA-like) noise levels is shown in red in Fig.~\ref{fig:forecast}. With better data (in this case more pulsars), we expect constraints of $C_\ell < 1$ up to multipole 6 using about 140 pulsars, with the dipole constraint reaching a level of a few percent. With 500 pulsars, reaching percent lever accuracy seems feasible.
In addition, the green curves in Fig.~\ref{fig:forecast} show the expected sensitivity for reduced white noise levels, as expected in the next generation of telescopes, what we dubbed the SKA-like scenario. In this case, constraining anisotropies at the level of 10\% becomes feasible for multipoles up to $\ell = 6$ for 140 pulsars, while 500 pulsars would achieve percent level accuracy.

We see that with the expected increase of high-quality data in the upcoming years, searching for anisotropies can become a powerful tool to distinguish possible sources of the observed GW signal. This new data will enable, or even mandate, further refinements in the analysis techniques. As an example, this will allow testing of more complex models for the frequency dependence of anisotropies. Besides the approach taken in this work, which is based on a factorization of the frequency and angular dependence of the GWB, other approaches based on modelling the frequency dependence expected from a population SMBHBs (see, e.g.,~\cite{Mingarelli:2013dsa,Becsy:2022pnr,Sato-Polito:2023spo,Lemke:2024cdu}) or more agnostic searches for anisotropies in different frequency bins independently (see, e.g.,~\cite{NANOGrav:2023tcn,Gersbach:2024hcc}) have been proposed. Once a first indication of an anisotropic signal is detected, these different methods will be crucial in determining the underlying physical model, and more data will allow for informative tests of more complex models. The FIM formalism presented in this work can be straightforwardly extended to these scenarios. 
Finally, we note that for a very significant increase in the number of pulsars, the assumption of dropping the oscillating pulsar term in Eq.~\eqref{eq:kappa} may need to be revisited for some pulsar pairs (located within a GW wavelength). The existence of such pulsar pairs is not expected to bias the existing analysis, however, additional information may be gained by retaining the pulsar term in these cases.

\subsubsection{Kinematic dipole}

One inevitable  deviation from isotropy is caused by the motion of the solar system barycenter (our local system) relative to the cosmological rest frame. Observation of the cosmic microwave background radiation
indicate that our local system is traveling at a velocity of $\beta_{\rm proper} = v/c =1.2 \cdot 10^{-3}$ towards the direction specified by galactic coordinates $(l = 264^\circ, b = 48^\circ)$. 
Assuming to have a similar velocity compared to the GWB rest frame, 
the anisotropies resulting from our proper motion can be described by
\be
\sum_{\ell m} c_{\ell m} Y_{\ell m} ( \theta, \phi)  = 1 +
\beta_{\rm proper} (\cos \theta \cos 97^\circ+ 
\sin \theta \sin 97^\circ \cos \phi) .
\ee
where the angles indicate spherical coordinates in the earth frame. %, while $\beta_{\rm proper} = 1.23 \times 10^{-3}$.
 One thus obtains the following multipole moments
\begin{align}
c_{00} &=1/ \sqrt{4 \pi} ;
\qquad
c_{1-1} = - c_{11} = \beta_{\rm proper} \sqrt{1 \over {6} } \sin 97^\circ ;
\qquad
c_{10} = \beta_{\rm proper} \sqrt{1 \over 3} \cos 97^\circ ;
% p_{1,1} = -\beta_{\rm proper} \sqrt{2 \pi \over 3} \sin 97^\circ 
\qquad
c_{\ell m} = 0 \quad {\rm for\ } \ell>1.
\end{align}
In the strong signal limit described in Sec.~\ref{sec:strongsignal}, we find that the $1\sigma$ uncertainty on 
$\beta_{\rm proper}$ scale as 
\begin{equation}
    \Delta \beta_{\rm proper} \simeq 0.8 \times 1/ \sqrt{N_p\, N_f}
    \qquad \quad
    (\text{strong signal limit}).
\end{equation}
This means that, in order to reach detection capability at around 90\% C.L., $N_f=10^2$ dominated by the signal for $N_p = {\cal O}(10^4)$ would be needed.
This forecast is similar, although more pessimistic, than what is reported in Ref.~\cite{Cruz:2024svc} (see also \cite{Tasinato:2023zcg}) due to their extrapolation of the weak-signal limit, which is not expected to hold within the future PTA configurations needed to reach such a level of precision.

\subsection{Cosmic Variance}
\label{sec:CV}

PTA experiments operate in a regime where neither the time average over observation time nor the volume average over the effective detector volume, set by the wavelength of the GWs, allows sampling many realizations of a GWB. Consequently, measurements may be sizeably affected by cosmic variance. This notably affects the measurement of the angular cross-correlation (HD curve), see Refs.~\cite{Allen:2022dzg,Bernardo:2022xzl,Allen:2022ksj,Allen:2024uqs}. In this subsection, we clarify the role of cosmic variance in measuring anisotropies. For simplicity, we will focus on the strong signal limit.

Our starting point for discussing anisotropies is the sky map of the power spectral density introduced in Eq.~\eqref{eq:hab_corr}. This is a well-defined physical quantity in the local universe, and measuring it is limited only by finite statistics and various noise contributions. This is consistent with the results shown in the right panel of Fig.~\ref{fig:Deltaclm}, where we show that the expected limits on anisotropy continuously decrease with larger numbers of pulsars in the strong signal limit.
However, we can only perform measurements in our local universe (given by effective detector volume and observation time). 
If this does not amount to sufficiently many independent measurements within the local universe (as is the case for PTAs), our observation may not be representative of the ensemble average found in the entire universe. In this sense, cosmic variance does not limit our ability to measure anisotropies in the local universe, but it does limit our ability to infer information about the parameters of the full ensemble average and hence the parameters of any underlying physics model.

\begin{figure*}[h]
    \centering
    \includegraphics[width=0.49\textwidth]{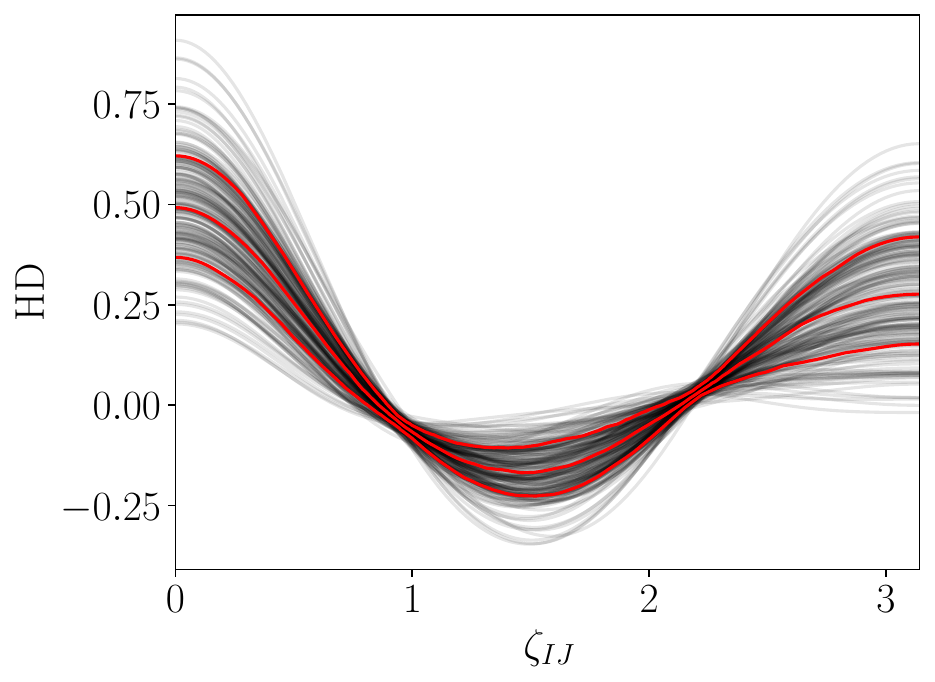}
    \includegraphics[width=0.49\textwidth]{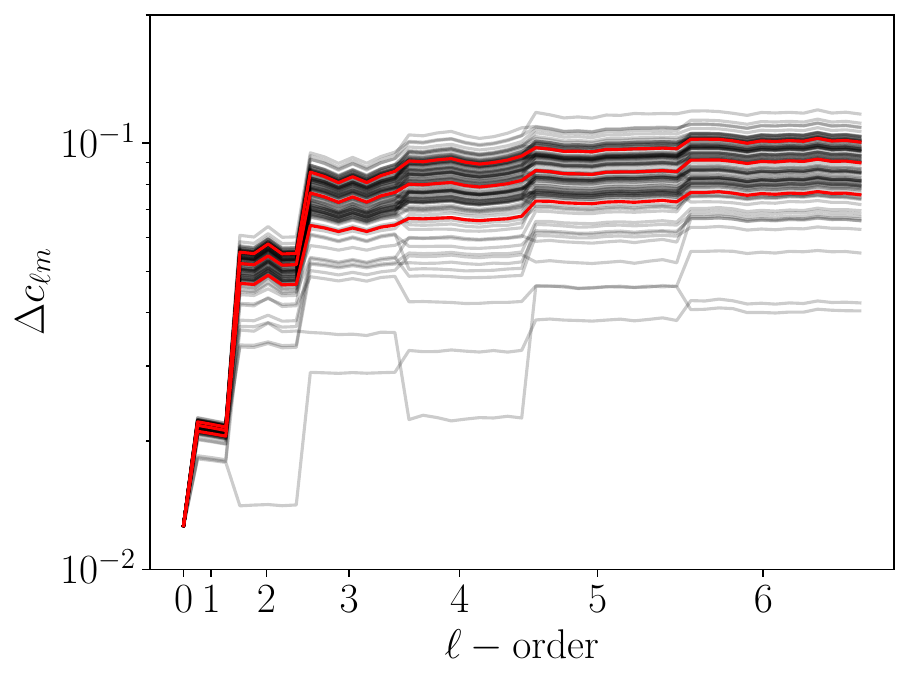}
    \caption{ {\it Left:} 100 realizations of $\chi_{IJ}$ including the effect of cosmic variance around the HD curve. In red we bracket the 68\% C.I. 
    {\it Right:} Uncertainties on $c_{\ell m}$ in the strong  signal limit with $N_p = 500$.
    As in the left panel, each black line corresponds to one of the 100 realizations of $\chi_{IJ}$, while the red curves indicate the corresponding $68\%$ C.I.}
    \label{fig:Deltaclm_cosmicvariance}
    \end{figure*}

There is another, more subtle impact of cosmic variance on the measurement of anisotropies. Note that the HD correlation enters our sensitivity estimates through the factors $\Gamma_{IJ,00}^{-1}$ in Eq.~\eqref{eq:strong-signal} as part of the PTA response function. In our sensitivity estimates, we used the mean value~\eqref{eq:HD_correlation} for the HD correlation. In any realization of the universe, cosmic variance implies slightly different realizations of the HD correlation~\cite{Allen:2022dzg}. This can be seen as a slight modification of the actual PTA response to anisotropies, resulting in slightly better or worse sensitivity. To illustrate this, Fig.~\ref{fig:Deltaclm_cosmicvariance} shows in the left panel different realizations of the HD correlation and on the right, the corresponding sensitivity to anisotropies. Here we have assumed an isotropic GWB and we are working in the strong signal limit. Overall, the impact is relatively small, at most 15\% at $1\sigma$. Remarkably, the dipole measurement is particularly insensitive to this cosmic variance effect ($\lesssim 3\%$ at $1\sigma$). 

\subsection{Hellings-Downs correlation}
It is well known that for an anisotropic GWB, the observed angular correlation function scatters around the HD function (see e.g.~\cite{Taylor:2013esa}). After averaging over many pulsar pairs, the HD function is recovered as the mean value\footnote{This implies taking an ensemble average, whereas any actual measurement will be limited by cosmic variance.} of the angular correlation function~\cite{Hotinli:2019tpc}, however with an increased variance reflecting the presence of anisotropies~\cite{Bernardo:2023jhs,Agarwal:2024hlj,Allen:2024bnk,Grimm:2024lfj,Allen:2024mtn}. This implies on the one hand, that the measurement of the HD correlation (averaged over many pulsars) is largely insensitive to the presence of anisotropies. On the other hand, this implies that the HD function is a robust criterion for a GW signal, both for isotropic GWBs, isolated sources, and anisotropic signals arising from multiple point sources~\cite{Cornish:2013aba,Cornish:2015ikx,Taylor:2020zpk,Becsy:2022pnr,Allen:2022dzg}. 

\begin{figure*}[h]
\centering
\includegraphics[width=0.32\textwidth]{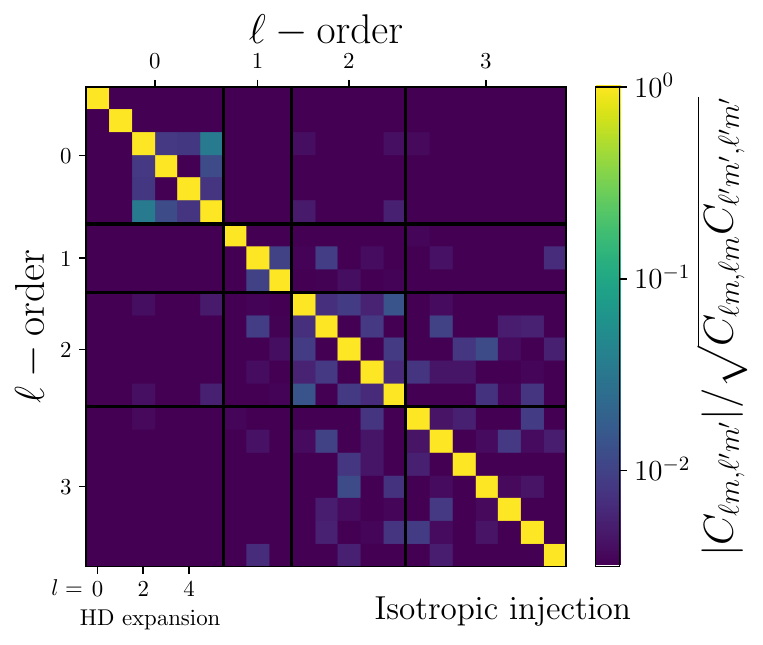}
\includegraphics[width=0.32\textwidth]{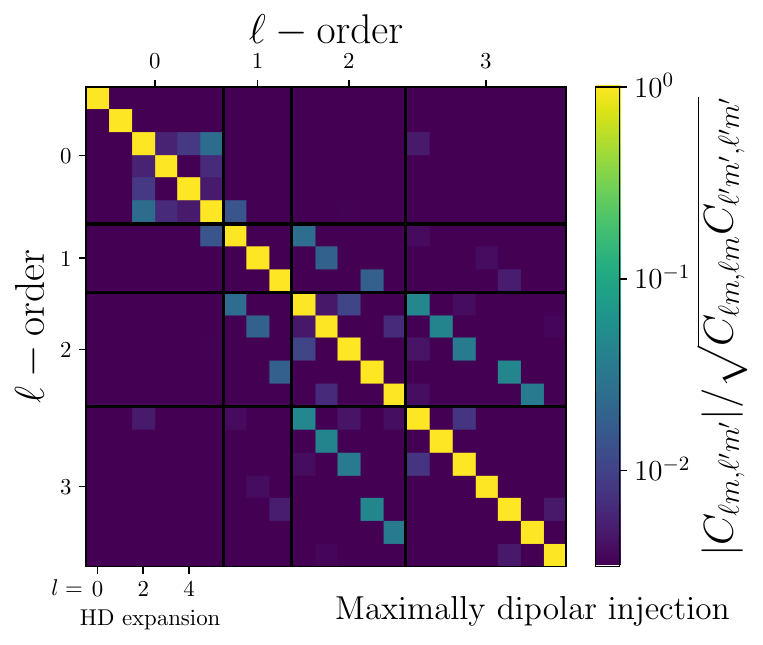}
\includegraphics[width=0.32\textwidth]{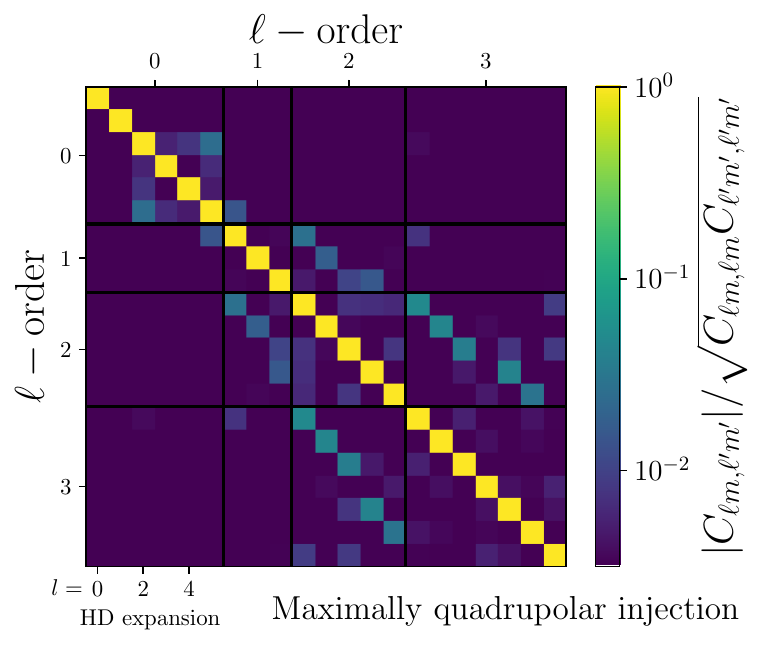}
\caption{Covariance matrix computed inverting the FIM in the noiseless limit. We include the HD expansion in Legendre polynomials $a_l$ with $l = 0,1, \dots, 5$ as well as anisotropy $c_{\ell m}$ coefficients for $\ell = 1,2,3$.
From left to right: isotropic injection $P(\hat k ) = 1/4 \pi$,  maximally dipolar injection $P(\hat k ) = (1+\cos\theta)/4 \pi$, and maximally quadrupolar injection $P(\hat k ) = 3\cos^2\theta/4 \pi$.
}
\label{fig:HD_anis_cov}
\end{figure*}

Here, we add to this story by remarking that the expansion of the HD function $\Gamma_{IJ,00}$ into Legendre polynomials (with coefficients $a_l$)\footnote{Following Ref.~\cite{Gair:2014rwa,Antoniadis:2023ott}, one can expand the HD function as $\Gamma_{IJ,00} = 
\chi_{IJ} 
=
\delta_{IJ}/2 + \sum_{l = 0}^{n} a_l P_l(\cos \zeta_{IJ}) $.
Using the standard normalization of the Legendre polynomials, the coefficients are found to be 
$a_l= 3 (2 l  +1) / 2 ( l +2)( l +1) l (l-1)$ 
for $l \geq 2$, and $a_0 =a_1 = 0$~\cite{Gair:2014rwa,Romano:2023zhb}.} and the expansion of the GWB in spherical harmonics (with coefficients $c_{\ell m}$) are to very good approximation uncorrelated in our measurements. To illustrate this, we show in Fig.~\ref{fig:HD_anis_cov} the magnitude of the entries of the covariance matrix in the strong signal limit when performing both of these expansions. We note that the resulting covariance matrix is, to good approximation, diagonal in the $(\ell,m)$-index and that in particular, the off-diagonal elements mixing the HD expansion with the GWB anisotropy coefficients are very small (and typically below few percent). This implies that the determination of the coefficients $a_l$ of the Legendre expansion is largely unaffected by the presence of anisotropies, as long as these are included in the model used to fit the data. We confirm this both for the injection of an isotropic signal (shown in Fig.~\ref{fig:HD_anis_cov} in the leftmost panel) as well as for the injection of maximally anisotropic signals, i.e., a maximally dipolar signal $P(\hat k ) = (1+\cos\theta)/4 \pi$ corresponding to 
an injection with only non-zero $\{c_{00} = 1/(2 \sqrt{\pi}), c_{10} = 1/(2 \sqrt{3 \pi})\}$ and a maximally quadrupolar signal $P(\hat k ) = 3\cos^2\theta/4 \pi$ corresponding to $\{c_{00} = 1/(2 \sqrt{\pi}),c_{20} = 1/(\sqrt{5 \pi})\}$.

\section{Conclusions}\label{sec:conclusions}

The search for anisotropies is likely going to be key in determining the origin of a nanohertz GW signal. In this paper, we perform forecasts using the FIM formalism to estimate the expected sensitivity of current and upcoming PTAs to anisotropies in a GWB at these frequencies. Our results are based on using the information from the full covariance matrix of pulsar pairs. We find that the geometric response of PTAs leads to a particularly good sensitivity to a dipole component in the GWB, while the sensitivity to higher multipoles is suppressed. As a consequence, current PTAs are not expected to obtain informative constraints on higher multipoles. Our results suggest that some constraints quoted in the literature, in particular in Ref.~\cite{NANOGrav:2023tcn}, appear to be dominated by the prior choice for all multipoles above the quadrupole. 

However, this situation will rapidly improve as PTA experiments increase the number of pulsars and reduce their noise levels. New pulsars are currently being discovered at a rate larger than 20/year~\cite{Verbiest:2024nid} thanks, especially to the new generation of highly sensitive telescopes FAST \cite{Hobbs:2014tqa} and MeerKAT \cite{Miles:2022lkg}. Along with SKA, future observations are expected to feature a white noise power spectrum reduced by an order of magnitude compared to current PTA measurements \cite{Weltman:2018zrl}.
This will significantly increase the sensitivity to anisotropies. For example, with the noise levels expected for SKA, the dipole sensitivity reaches 1\% and all multipoles up to $\ell = 6$ can be constrained to better than 10\% using 100 pulsars. A further order of magnitude improvement seems possible with about 500 pulsars.
Beyond increasing the overall sensitivity, reduced white noise levels will increase the frequency range of PTAs to higher frequencies, which are currently white noise dominated. This is particularly interesting for the prospect of searching for anisotropies expected in a GWB from SMBHBs, which increase at higher frequencies due to the smaller number of binaries per frequency bin~\cite{NANOGrav:2023tcn}.

A further interesting outcome of our analysis is the remarkably strong sensitivity to dipolar anisotropies. This effect can be understood based on purely geometric observations in the strong signal limit and can be traced back to the approximately diagonal structure of the generalized dipole overlap reduction function in the pulsar indices. Somewhat surprisingly, this feature is not reflected in the limits reported by the NANOGrav collaboration in their recent Bayesian anisotropy  search~\cite{NANOGrav:2023tcn}. We propose several possible explanations for this discrepancy, keeping in mind that our results are based on Fisher forecasts as opposed to actual data, as well as on a simplified pulsar noise model. We expect upcoming data releases, which will significantly improve the ability to search for anisotropies, to shed more light on this question.

At the repository linked in Ref.~\cite{code_repo}, we include the code \href{https://github.com/Mauropieroni/fastPTA/}{\texttt{fastPTA}} readily usable to estimate the sensitivities and the measurement uncertainties with current and future PTA configurations.

\begin{acknowledgments}
We thank Nihan Pol for the useful discussion and clarifications on the NANOGrav search for anisotropies and Fabrizio Rompineve for collaboration on the initial stages of the project. We thank Bruce Allen for valuable discussions on the HD cosmic variance and in-depth comments on the draft. Moreover, we thank Stanislav Babak, and Carlo Contaldi for the insightful discussions. We thank Ben Farr and Andrea Mitridate for comments on the draft. MP thanks Rutger van Haasteren for the useful discussion on PTA data analysis. MP acknowledges the hospitality of Imperial College London, which provided office space during some parts of this project.
\end{acknowledgments}

\appendix

\section{Approximations in the derivation of the covariance matrix in the frequency domain}
\label{app:cov_dft}

The derivation of the covariance matrix in frequency space, Eq.~\eqref{eq:cov_fourier_approx} in the main text, relies on several approximations, which we critically revisit in this appendix. In particular, we investigate the impact of the unknown GW phase and possible correlations between different frequencies\footnote{For the impact of correlations between frequency bins on the determination of the HD curve see~\cite{Allen:2024uqs}. }. Since these questions do not depend on the angular properties of the signal, for simplicity, we restrict ourselves to the isotropic case. Similarly, we will also neglect the noise terms and deterministic features induced by errors of pulsar timing model parameters.

In the following, we assume that there are $N$ observations at $t_j = -T_\text{obs}/2 + j T_\text{obs} / N$, $j = 0, \dots, N-1$, of all $\Delta t_I$ with pulsar index $I$ from 0 to $N_p$.
Eq.~\eqref{eq:tim_res_gen} in the main text gives the time delay induced by a GW compared to propagation of the pulsar signal in an Minkowski background. At frequencies $f \gg 1/T_\text{obs}$, fitting the timing model to the full dataset achieves this by effectively averaging over many GW oscillations. On the contrary, at low frequencies, $f \sim 1/T_\text{obs}$, one is left with a residual sensitivity to the unknown GW phase at $t = 0$. To illustrate this, we assume a timing model to be calibrated to $\Delta t (t = 0) = 0$, leading to
\begin{equation}
    \Delta t _I(t) =
    \frac{1}{2}
    \frac{\hat{p}_I^a \hat{p}_I^b}{(1+\hat{p}_I\cdot \hat k)}
    \int_0^t {\rm d} t'
    \,
\left[ h_{ab} \left( t', \vec{0}\right) - h_{ab}  \left( t'-D_I, \hat{p}_I D_I \right)\right]
     \label{eq:tim_res_app} \,.
\end{equation}
From this, we can compute the covariance matrix following the sthe onset of the observation (ame steps as in Sec.~\ref{sec:Response_ORF} of the main text,
\begin{align}
 C_{t, h, IJ, ij} &= \langle \Delta t_I (t_i) \Delta t_J (t_j) \rangle  = \chi_{IJ} \int_{-\infty}^\infty {\rm d}f \frac{S_h(f)}{24 \pi^2 f^2} \left( e^{2 \pi i f t_i} - 1 \right)  \left( e^{-2 \pi i f t_j} - 1 \right)  \,,\label{eq:C_h_time_app_aux}
\end{align}
where $\chi_{IJ}$ is the HD correlation reported in Eq.~\eqref{eq:HD_correlation}.
Compared to Eq.~\eqref{eq:C_h_time} in the main text, we note the extra `-1' terms arising from the `miscalibration' of the pulsar timing model. Upon moving to the frequency domain, and assuming $S_h(f)/f^2$ to be approximately constant, these terms will contribute only to the $f_i = 0$ mode (as expected from their role as free integration constant in Eq.~\eqref{eq:tim_res_app}) and thus not impact the analysis within the PTA band (see also related discussion in Ref.~\cite{vanHaasteren:2012hj}). We thus proceed with the expression for the covariance matrix used in the main text,
%As discussed in Sec.~\ref{sec:Response_ORF}, the complete covariance matrix is then constructed from
\begin{align}
C_{t, h, IJ, ij} &= \langle \Delta t_I (t_i) \Delta t_J (t_j) \rangle \nonumber = \chi_{IJ} \int_{-\infty}^\infty {\rm d}f \frac{S_h(f)}{24 \pi^2 f^2} e^{2 \pi i f (t_i - t_j)}  \,. \label{eq:C_h_time_app}
\end{align}
%where $\chi_{IJ}$ is the HD correlation reported in Eq.~\eqref{eq:HD_correlation}.
In the case of anisotropies in the GWB, this correlation must be replaced by the overlap reduction function $\Gamma_{IJ}$ given in Eq.~\eqref{eq:ORF}. Defining the vector of all timing residuals 
\begin{align}
\Delta t = \begin{pmatrix}
    \Delta t_0 (t_0) \\
    \vdots \\
    \Delta t_0 (t_{N-1}) \\
    \Delta t_1 (t_0) \\
    \vdots \\
    \Delta t_{N_p} (t_{N-1})
\end{pmatrix} = \begin{pmatrix}
    \Delta t_0 \\
    \vdots \\
    \Delta t_{N_p}
\end{pmatrix} \, ,
\end{align}
and organising the complete covariance matrix $C_h$ with entries given by Eq.~\eqref{eq:C_h_time_app} accordingly, the Gaussian probability density of observing timing residuals $\Delta t$ is given by
\begin{align}
p (\Delta t | C_{t,h}) = \frac{1}{\sqrt{(2 \pi)^{N N_p} \det C_{t,h}}} \exp \left( -\frac{1}{2} \Delta t^T C_{t,h}^{-1} \Delta t \right)\,.
\end{align}
With the transformation matrix
\begin{align}
U_I = \frac{1}{\sqrt{N}} \begin{pmatrix}
    e^{i 2 \pi f_{k_\mathrm{min}} t_0} & \dots & e^{i 2 \pi f_{k_\mathrm{max}} t_0} \\
    \vdots & \ddots & \vdots \\
    e^{i 2 \pi f_{k_\mathrm{min}} t_{N-1}} & \dots & e^{i 2 \pi f_{k_\mathrm{max}} t_{N-1}}
\end{pmatrix} \, ,
\end{align}
we can express the timing residuals $\Delta t_I$ via the discrete Fourier transform\footnote{We include the prefactor $\sqrt{N}/T_\text{obs}$ to match the dimensions of~\cite{Babak:2024yhu}.}
\begin{align}
\Delta t_I = \frac{\sqrt{N}}{T_\text{obs}} U_I \widetilde{\Delta t}_I\,,
\end{align}
where the elements of $\widetilde{\Delta t}_I$ correspond to the Fourier coefficients with frequencies $f_k = k/T_\text{obs}$, where $k = k_\mathrm{min}, k_\mathrm{min}+1, \dots, k_\mathrm{max}$, $k_\mathrm{min} = -(N-1)/2$ and $k_\mathrm{max} = (N-1)/2$ for odd $N$, $k_\mathrm{min} = -N/2 + 1$ and $k_\mathrm{max} = N/2$ for even $N$. We note that $U_I^\dag = U_I^{-1}$. Combining all transformation matrices into
\begin{align}
U = \begin{pmatrix}
    U_0 \\
    & \ddots \\
    & & U_{N_p-1}
\end{pmatrix} \,,
\end{align}
we find the probability density in terms of the Fourier coefficients, again combined into a single vector,
\begin{align}
p (\Delta t | C_{t,h}) &= \frac{1}{\sqrt{(2\pi)^{N N_p} \det ( C_{t,h})}} \exp \left( -\frac{N}{2 T_\text{obs}^2} \widetilde{\Delta t}^\dag U^\dag C_{t,h}^{-1} U \widetilde{\Delta t} \right) = \frac{1}{\sqrt{(2 \pi N / T_\text{obs})^{N N_p} \det C_h}} \exp \left( -\frac{1}{2} \widetilde{\Delta t}^\dag C_h^{-1} \widetilde{\Delta t} \right)\,.
\end{align}
This probability distribution for the Fourier coefficients resembles a complex normal distribution with covariance matrix $C_h = ( T_\text{obs}^2/N) U^\dag C_{t,h} U$. As the timing residuals are real numbers there are however additional conditions $\Im \widetilde{\Delta t}_{I,k=0} = \Im \widetilde{\Delta t}_{I,k=N/2} = 0$ and $\widetilde{\Delta t}_{I,-k} = \widetilde{\Delta t}_{I,k}^*$. The components of the covariance matrix are given by
\begin{align}
C_{h,IJ,kl} &= \frac{T_\text{obs}^2}{N^2} \sum_{m, n} e^{-i 2 \pi (f_k t_m - f_l t_n)} \langle \Delta t_{I} (t_m) \Delta t_J (t_n) \rangle \\
&= \frac{T_\text{obs}^2}{N^2} \chi_{IJ} \int_{-\infty}^\infty {\rm d} f \frac{S_h(f)}{24 \pi^2 f^2} \;  \frac{e^{-i \pi T_\text{obs} (f - f_k)} - e^{i \pi T_\text{obs} (f - f_k)}}{1 - e^{i 2 \pi T_\text{obs} (f - f_k) / N}}  \; \frac{e^{i \pi T_\text{obs} (f - f_l)} - e^{-i \pi T_\text{obs} (f - f_l)}}{1 - e^{-i 2 \pi T_\text{obs} (f - f_l) / N}} \,. 
\end{align}
%In the low-frequency regime with large $N$, i.e., for $T_\text{obs} (f - f_k) / N \ll 1$ and $T_\text{obs} (f - f_l) / N \ll 1$, we can approximate
In PTAs, observation time and pulsar white noise limit the observation band to $f_\text{min} \sim 1/T_\text{obs} < f < f_\text{max}$ with $f_\text{max} \sim 10 f_\text{min} \ll N f_\text{min}$. Consequently, we can approximate  $T_\text{obs} (f - f_k) / N \ll 1$ to obtain
\begin{align}
C_{h,IJ, kl} &\simeq T_\text{obs}^2 \chi_{IJ} \int_{-\infty}^\infty {\rm d} f \frac{S_h(f)}{24 \pi^2 f^2} \mathrm{sinc} \left[ \pi T_\text{obs} (f - f_k)  \right]  \mathrm{sinc} \left[ \pi T_\text{obs} (f - f_l) \right] \label{eq:cov_fourier}  \\
& \simeq T_\text{obs} \, \chi_{IJ} \,  \frac{S_h(f)}{24 \pi^2 f^2} \mathrm{sinc}\left[ (f_k - f_l) \pi T_\text{obs}\right] \\
& \simeq \chi_{IJ} \frac{S_h (f_k)}{24 \pi^2 f_k^2} \delta(f_{k} -f_{l})\,. \label{eq:cov_fourier_approx_app_aux}
\end{align}
In the second step, we have assumed $S_h(f)/f^2$ to be approximately constant (otherwise analogous results can be obtained after performing an appropriate whitening procedure). For the third step, recall that $f_k$ take integer multiples of $1/T_\text{obs}$, which are zeros of the sinc function for any $k \neq l$.
% we have made use of the sinc representation of the Dirac delta function in the limit $f_{\{k,l\}} T_\text{obs} \gg 1$. With $f_\text{min} \sim 1/T_\text{obs}$ this is a priori not a particularly good approximation in the low-frequency regime of PTAs, and we investigate the impact of corrections to this below. 
Finally, given that Eq.~\eqref{eq:cov_fourier_approx_app_aux} is symmetric in positive and negative frequencies, we can restrict ourselves to positive frequencies only and obtain
%Restricting to positive frequencies,  $k, l > 0$ and following the same steps between \eqref{eq:Chfinal} and \eqref{eq:cov_fourier_approx}, one finds
% and if $S_h (f) / f^2$ is approximately constant for frequencies $f - f_k \lesssim 1/T_\text{obs}$ and $f - f_l \lesssim 1/T_\text{obs}$ one may approximate $\mathrm{sinc} (\pi T_\text{obs} [f-f_k]) \mathrm{sinc} (\pi T_\text{obs} [f-f_l]) \simeq 2 \delta_{kl} \delta (f - f_k) / T_\text{obs}$ leading to the expression used in the main text,
\begin{align}
C_{h,IJ, kl}|_{k,l > 0} &\simeq \chi_{IJ} \frac{S_h (f_k)}{12 \pi^2 f_k^2} \delta(f_{k} -f_{l})\,. \label{eq:cov_fourier_approx_app}
\end{align}

\begin{figure*}[t]
    \centering
    \includegraphics[width=0.467\textwidth]{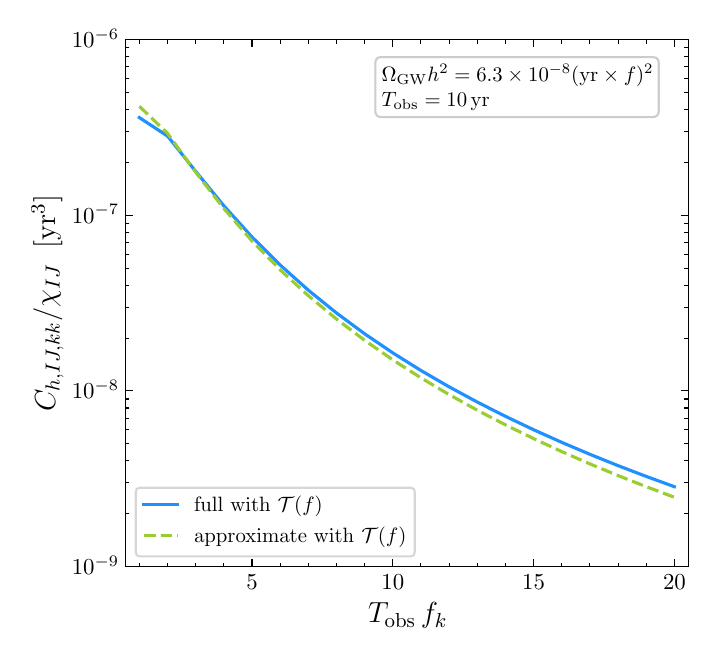}
    \includegraphics[width=0.513\textwidth]{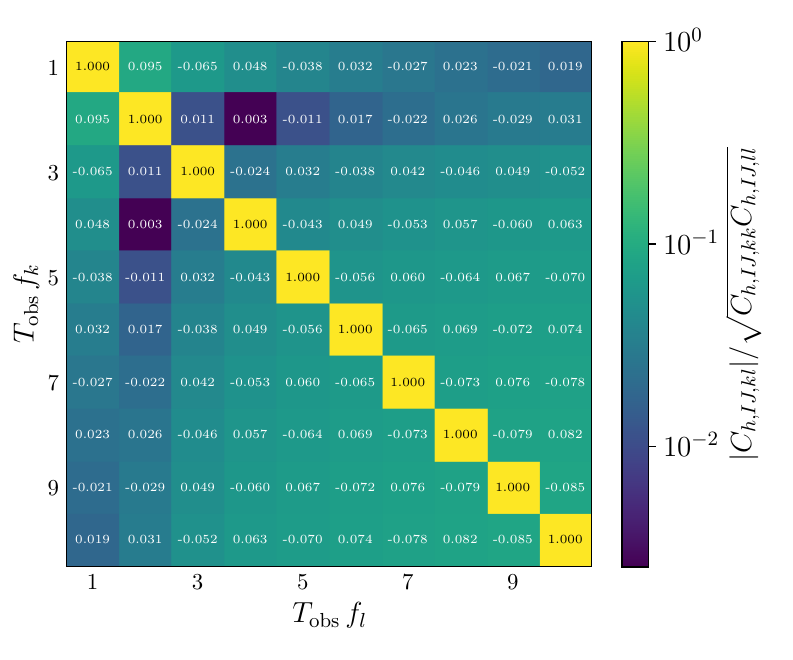}
    \caption{
    \textit{Left:} Diagonal part of covariance matrix for Fourier coefficients $f_k$ from the full expression in Eq.~\eqref{eq:cov_fourier} (blue full) as well as the approximate Eq.~\eqref{eq:cov_fourier_approx_app}. We divide by the HD correlation and assume 10 yr observations with $\Omega_\mathrm{GW} h^2 = 6.3 \times 10^{-8} (\mathrm{yr} \times f)^2$~\cite{Antoniadis:2023ott} with the transmission function $\mathcal{T}(f)$ included, i.e., replacing $S_h \to \mathcal{T} S_h$ in Eqs.~\eqref{eq:cov_fourier} and~\eqref{eq:cov_fourier_approx_app}. \textit{Right:} Covariance matrix with diagonal factored out for the first 10 positive frequencies. Note that we use the absolute values for the color scale, but include the sign when writing explicit values. }
    \label{fig:cov_fourier}
    \end{figure*}

In the left panel of Fig.~\ref{fig:cov_fourier} we show the diagonal part of the covariance matrix for Fourier coefficients in Eq.~\eqref{eq:cov_fourier} (blue full) compared to the approximation in Eq.~\eqref{eq:cov_fourier_approx_app} (green dashed). Here, we divide by the HD correlation and assume a signal consistent with current PTA observations $\Omega_\mathrm{GW} h^2 = 6.3 \times 10^{-8} (\mathrm{yr} \times f)^2$~\cite{Antoniadis:2023ott}. We have moreover generalized the expressions in Eq.~\eqref{eq:cov_fourier} and in Eq.~\eqref{eq:cov_fourier_approx} by replacing $S_h \to \mathcal{T} S_h$ to include the
transmission function $\mathcal{T}(f) = [1 + 1/(f T_\text{obs})]^6$~\cite{Babak:2024yhu,Hazboun:2019vhv} to model the loss in sensitivity at low frequency and make the covariance matrix finite for the assumed spectrum. The transmission function causes the slight dip at $T_\text{obs} f_k = 1$. The approximation differs from the full result by $\sim 1 - 16 \%$ for non-zero frequencies. In the right panel of Fig.~\ref{fig:cov_fourier} we show the full covariance matrix for the first 10 non-zero frequencies. We clearly see that the diagonal dominates, but the off-diagonal components can reach $1 - 10\%$. This indicates that the approximation taking the covariance matrix to be diagonal in frequency with entries as given in Eq.~\eqref{eq:cov_fourier_approx} is valid at the few-percent level, but deserves further investigation in the future with increased PTA sensitivity. Note however that for less smooth GWB spectra than the power law we assume here, the difference between the approximation and the full result can generally be much larger. See also~\cite{Allen:2024uqs,Allen:2024bnk} for a discussion on the correlations between frequency bins. 

As a final comment, we stress that in the strong signal limit discussed in~\cref{sec:strongsignal}, even if present, correlations between frequencies do not impact the determination of the $c_{\ell m}$ coefficients. This follows from the signal and FIM structure in the frequency and $\ell m$ spaces. Since the $c_{\ell m}$ coefficients are frequency-independent, the $M$ matrix appearing in Eq.~\eqref{eq:FIM_M_mat} multiplies the identity in the frequency space and the $N_f$ scaling of the FIM is unaffected.

\section{Generalised overlap reduction functions}\label{appendix:ORF}

In this appendix, we report the analytical expressions for the generalized ORFs in the so-called ``computational frame''~\cite{Mingarelli:2013dsa, Gair:2014rwa, Kato:2015bye, AnilKumar:2023yfw, Allen:2024bnk}. This corresponds to aligning the $\hat z$ axis with the first pulsar direction $\hat p_I =(0,0,1)$, while the second pulsar is placed in the $(\hat x, \hat z)$ plane, i.e., 
$\hat p_J=(\sin\zeta,0,\cos\zeta)$. 
Higher multipoles can be found at~\cite{Gair:2014rwa, Kato:2015bye, AnilKumar:2023yfw}.

The expressions reported here are obtained by expanding in complex spherical harmonics,
\begin{equation}
    \tilde \Gamma_{IJ,\ell m} 
    \equiv 
    \kappa_{IJ}
    \int 
    {\rm d}^2 \Omega_{\hat k} 
    \gamma_{IJ}(\hat k) 
    {\cal Y}_{\ell m}(\hat k) \,,
    \label{eq:GammaIJlm}
\end{equation}
and can be recast to obtain the functions $\Gamma_{IJ,\ell m}$ in the basis of real spherical harmonics used in the main text using Eq.~\eqref{eq:realYlm}, to get
\beq
\Gamma_{IJ,lm} =  
\left\{
\begin{array}{lr}
\frac{1}{\sqrt{2}} \left[\tilde \Gamma_{IJ,lm} 
+ (-1)^m \,\tilde \Gamma_{IJ,l-m}\right]		&
\quad \quad 	m > 0
\\
\tilde \Gamma_{IJ,l0} & 	\quad\quad 	m = 0 
\\
\frac{1}{i\sqrt{2}} \left[\tilde \Gamma_{IJ,l-m} 
- (-1)^m \,\tilde \Gamma_{IJ,lm}
\right] 	& 	\quad\quad 	m < 0\\
\end{array}
\right. \,.
\label{e:abGammalm_real}
\eeq

Given an array of pulsars in the sky, the ORFs are uniquely defined, and here expressed in terms of the pulsar index $I,J$ as well as the multipole indices $\ell,m$.
One can rotate from the computational frame to the cosmic rest frame using the three rotation angles
\begin{align}
\alpha & = \phi_I \,,
\qquad
\beta = \theta_I\,,
\qquad 
\tan \gamma  = \frac{\sin\theta_J \sin (\phi_J - \phi_I)}{\cos\theta_I \sin\theta_J \cos ( \phi_I - \phi_J)  - \sin\theta_I \cos\theta_J}\,,
\end{align}
with $(\theta_I,\phi_I$) and $(\theta_J,\phi_J$) denoting the direction of the pulsars $I$ and $J$, respectively. Then, 
\beq
\tilde  \Gamma_{IJ,\ell m} = 
\sum_{k=-\ell}^{\ell} [D_{\ell m k}(\alpha,\beta,\gamma)] \Gamma'_{IJ,\ell k},
\label{e:abGammalm_rot}
\eeq
with $D_{\ell m k}(\alpha,\beta,\gamma)$ being explicitly given in Eq.~(52) of~\cite{Mingarelli:2013dsa}. As in \cite{Mingarelli:2013dsa}, we denote the generalized (complex) ORFs in the computational frame by $\Gamma'_{IJ,\ell m}$.

\subsection{Explicit expressions for monopole, dipole, and quadrupole.}
\begin{figure*}[t]
    \centering
    \includegraphics[width=0.325\textwidth]{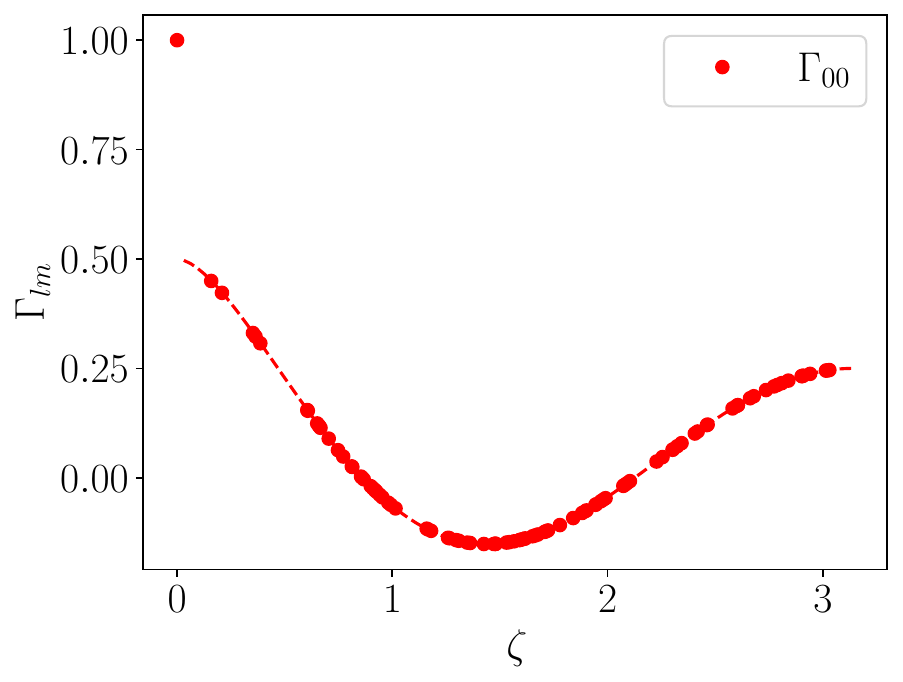}
    \includegraphics[width=0.325\textwidth]{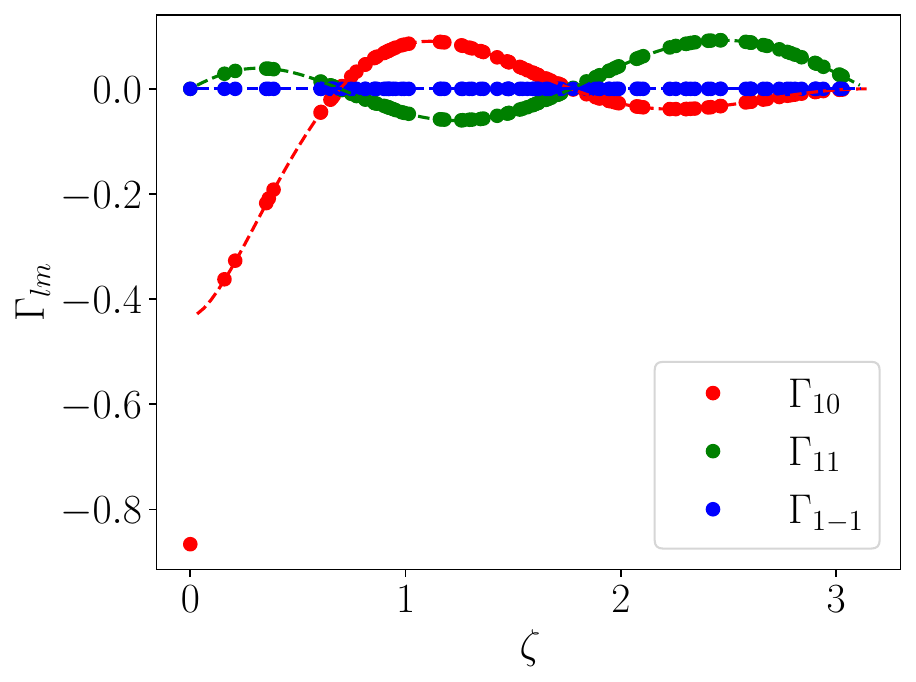}
    \includegraphics[width=0.325\textwidth]{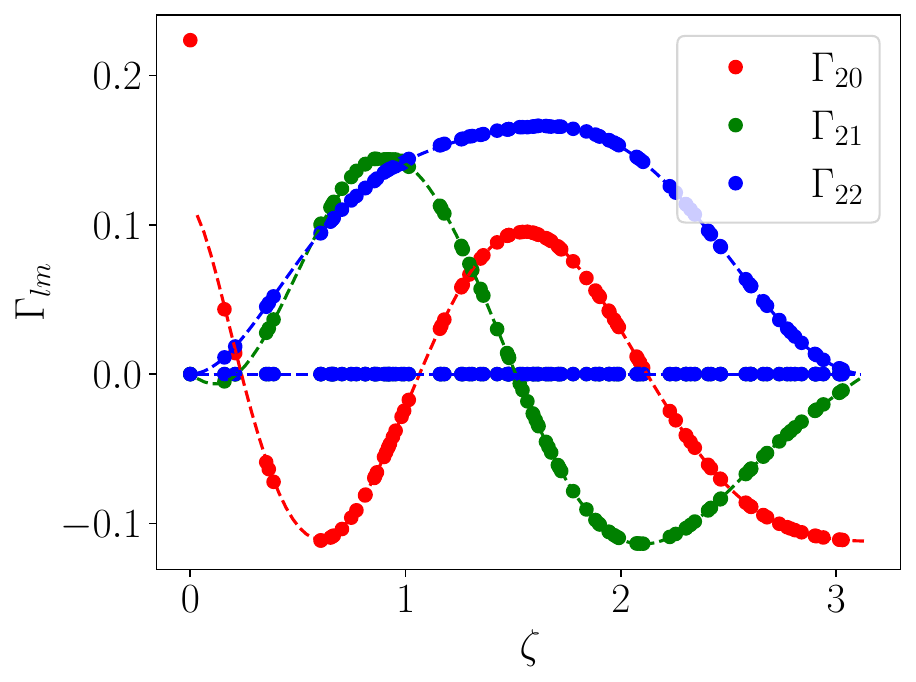}
    \caption{ Real-valued ORF in the computational frame as a function of the pulsar-pulsar angular separation $\zeta$. We include pulsar auto-correlations. The dots report the ORFs between a pulsar located along $\hat z$ and a random set of 100 pulsars in the $(\hat x, \hat z)$ plane, computed numerically by projecting Eq.~\eqref{eq:gamma} on the spherical harmonics. The dashed lines indicate the analytical results in Eqs.~\eqref{e:HD}, \eqref{e:gamma_dipole}, and \eqref{e:gamma_quadrupole}.}
    \label{fig:Gamma_real}
    \end{figure*}

For an isotropic background only the coefficient $\ell =m = 0$ is non-vanishing. One finds
\begin{equation}
\Gamma'_{IJ,00}=\frac{3}{8}
\left[1+\frac{\cos\zeta}{3}+4(1-\cos\zeta)\ln\left(\sin\frac{\zeta}{2}\right)\right](1+\delta_{IJ}),
\label{e:HD}    
\end{equation}
which is the HD curve~\cite{Hellings:1983fr}.
The corresponding expressions for the dipole are given by
\begin{subequations}
\label{e:gamma_dipole}
\begin{align}
\Gamma'_{IJ,1-1}&=-
\frac{1}{16}\sin\zeta\left\{1+3 (1-\cos\zeta) \left[1+\frac{4}{(1+\cos\zeta)}\ln\left(\sin\frac{\zeta}{2}\right)\right] \right\}(1+\delta_{IJ})\,,
\\
\Gamma'_{IJ,10}&=-
\frac{\sqrt{3}}{8}
\left\{(1+\cos\zeta)+3 (1-\cos\zeta) \left[(1+\cos\zeta)+4\ln\left(\sin\frac{\zeta}{2}\right)\right]\right\}(1+\delta_{IJ})\,,
\\
\tilde\Gamma_{IJ,11}&= - \Gamma_{IJ,1-1}\,.
\end{align}
\end{subequations}
and for the quadrupole one finds
\begin{subequations}
\label{e:gamma_quadrupole}
\begin{align}
&\Gamma'_{IJ,2-2}=\Gamma_{IJ,22}\,,\nonumber\\
&\Gamma'_{IJ,2-1}=-\Gamma_{IJ,21}\,,
\nonumber\\
&\Gamma'_{IJ,20}=
\frac{1}{4 \sqrt{5}}
\left\{\cos\zeta\!+\!\frac{15}{4} (1-\cos\zeta)\left[(1+\cos\zeta)(\cos\zeta\!+\!3)+\!8\ln\!\!\left(\sin\frac{\zeta}{2}\right)\right]\!\right\}(1+\delta_{IJ})\,,
\\
&\Gamma'_{IJ,21}=
\frac{1}{8}\sqrt{\frac{3}{10}}
\sin\zeta\!\left\{5\cos^2\zeta\!+\!15\cos\zeta\!-\!21\!-\!60\frac{(1-\cos\zeta)}{(1+\cos\zeta)}\ln\!\!\left(\sin\frac{\zeta}{2}\right)\right\}(1+\delta_{IJ})\,,
\\
&
\Gamma'_{IJ,22}
=
-
\frac{1}{16}\sqrt{\frac{15}{2}}
\frac{(1-\cos\zeta)}{(1+\cos\zeta)}\left[(1+\cos\zeta)(\cos^2\zeta\!+\!4\cos\zeta-9)-24 (1-\cos\zeta)\ln\!\!\left(\sin\frac{\zeta}{2}\right) \right](1+\delta_{IJ})\,.
\end{align}
\end{subequations}
These expressions are visualized in Fig.~\ref{fig:Gamma_real} (and similar figures are shown in~\cite{Mingarelli:2013dsa}). We note in particular the sizeable hierarchy between the values at $\zeta = 0$ and $\zeta \neq 0$ for the monopole and dipole, which leads to a particularly good sensitivity of PTAs to these low multipoles.

\section{Impact of increased noise levels and injected anisotropies on sensitivity estimates}

In this appendix, we present some additional results derived assuming different noise contributions or anisotropic injections, complementing the discussion in the main text. 

\subsection{Estimates for current sensitivity}

As discussed in the main text, our Fisher forecast for a PTA configuration based on the locations of pulsars in the NG15 data release results in somewhat better sensitivity to the dipole compared to the limits reported in Ref.~\cite{NANOGrav:2023tcn}. To explore the possible origins of this discrepancy, in this appendix, we show variations of our analysis. 

\begin{figure*}[h!]
\centering
\includegraphics[width=0.49\textwidth]{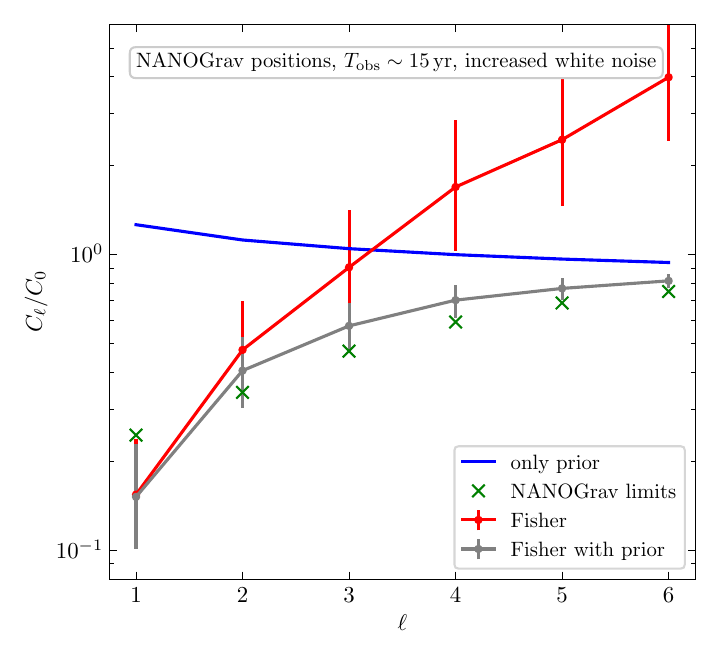}
\includegraphics[width=0.49\textwidth]{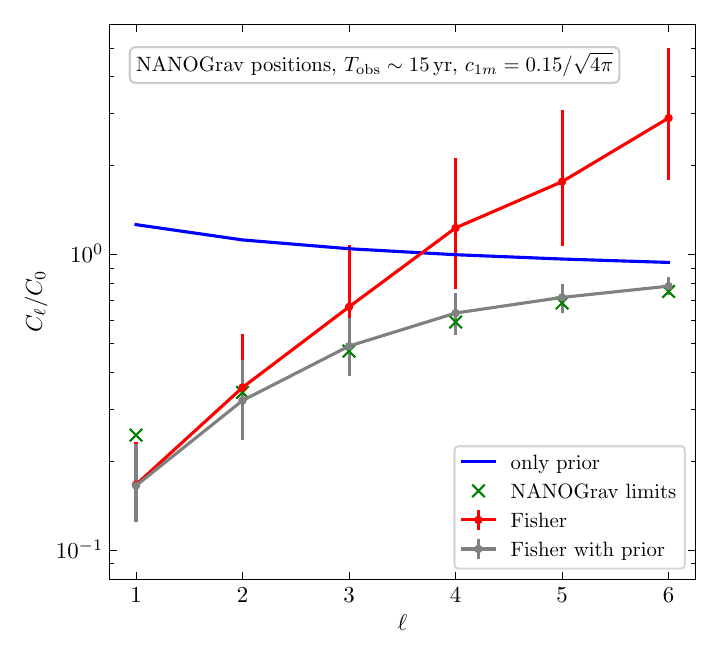}
\caption{
Same as left panel of Fig.~\ref{fig:ng_comparison}, but with power spectrum for white noise increased by 5\% (left) or dipole with $c_{1m} = 0.15/\sqrt{4 \pi}$, i.e., $C_1/C_0 = 0.0225$, injected (right).}
\label{fig:ng_comparison2}
\end{figure*}

Our analysis in the main text is based on assuming EPTA-like noise is uniformly sampled from the same distribution for all pulsars. In the left panel of Fig.~\ref{fig:ng_comparison2} we show results obtained when increasing the white noise power spectrum by $5\%$ in all pulsars. This agrees with the upper limit reported for the dipole in Ref.~\cite{NANOGrav:2023tcn} at 95\% CL, at the cost of a (mild) tension at higher multipoles.
Alternatively, we consider the impact of a dipole anisotropy, modeled here by taking the mean value for the injected $c_{1m}$ coefficients to be non-zero. As demonstrated in the right panel of Fig.~\ref{fig:ng_comparison2} a dipole with magnitude $C_1/C_0 \simeq 2\%$ would result in an upper bound comparable with the results reported in Ref.~\cite{NANOGrav:2023tcn}. Anisotropic noise components could mimic a similar effect.

\subsection{Forecast for scaling of future sensitivity in the strong signal limit}\label{app:anisotropic injection SSL}

\begin{figure*}[h!]
\centering
\includegraphics[width=0.49\textwidth]{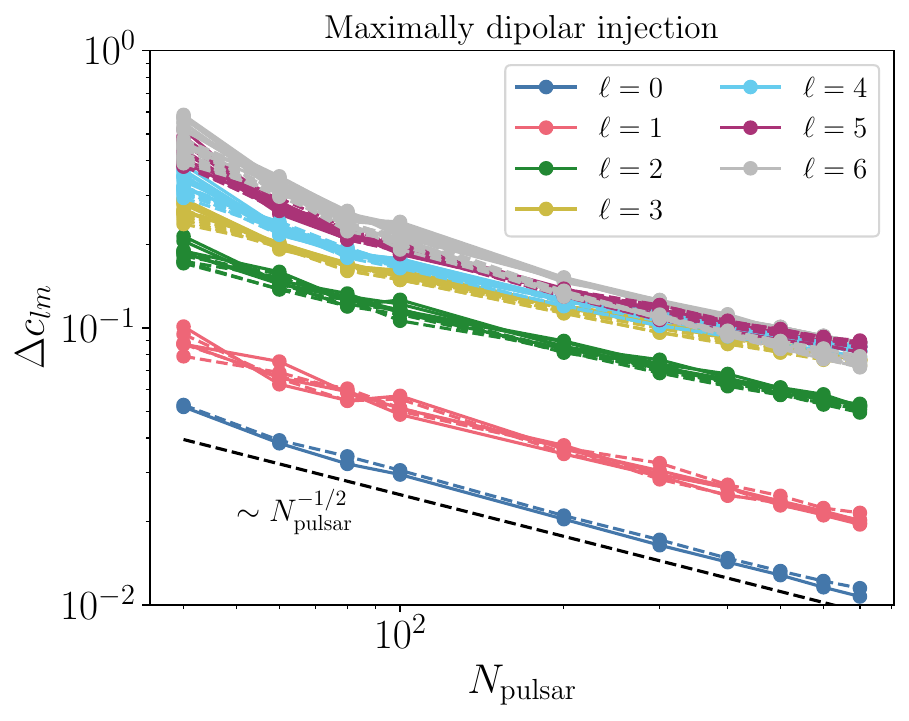}
\includegraphics[width=0.49\textwidth]{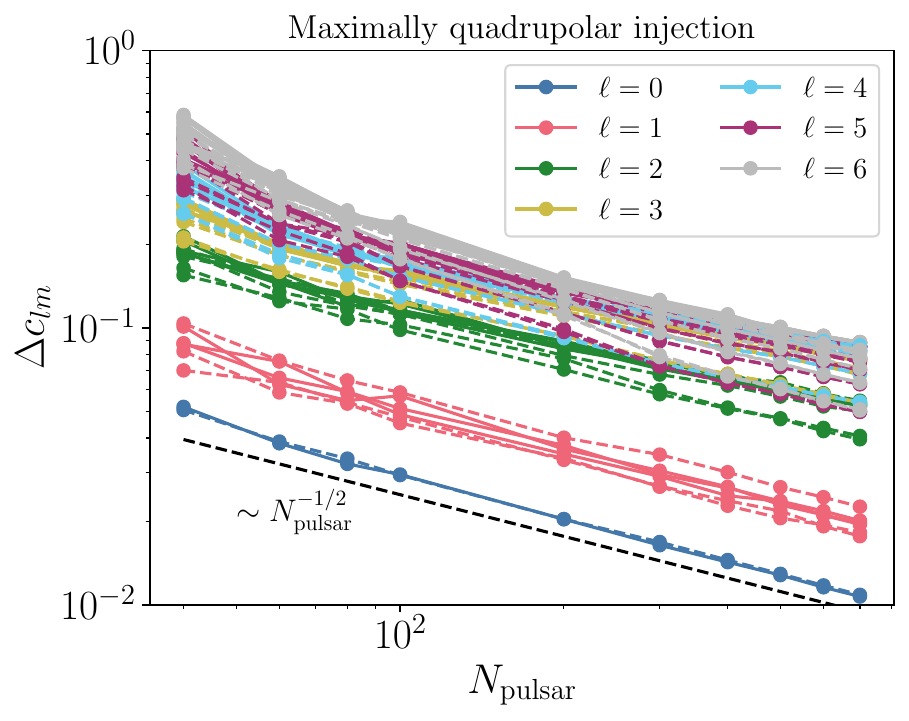}
\caption{
Same as Fig.~\ref{fig:Deltaclm} but for the case of a maximally dipolar injection $\{c_{00} = 1/(2 \sqrt{\pi}), c_{10} = 1/(2 \sqrt{3 \pi})\}$ (left panel) and maximally quadrupolar injection $\{c_{00} = 1/(2 \sqrt{\pi}),c_{20} = 1/(\sqrt{5 \pi})\}$ (right panel). 
The dashed lines indicate the results with injected anisotropies, while with solid lines assume isotropic injection (i.e., Fig.~\ref{fig:Deltaclm}, right panel) for comparison. 
}
\label{fig:Deltaclm_anisotropies}
\end{figure*}

In Fig.~\ref{fig:Deltaclm_anisotropies} we show the evolution of the uncertainty $\Delta c_{\ell m}$ as a function of the number of pulsars in the strong signal limit and assuming an anisotropic injection. In the left panel, we assume a maximally dipolar GWB, while in the right panel, the injected signal is maximally quadrupolar. 
These results extend the one presented in Fig.~\ref{fig:Deltaclm} and support the interpretation presented in the main text.

In particular, we observe that the uncertainties mostly scale as in the case of an isotropic injection, i.e. $\sim N_p^{-1/2}$ for $\ell =0,1$ and with a slightly milder scaling for larger $\ell$. 
We also see the appearance of differences between different $m$, due to the introduction of a preferred direction associated with the maximal injected anisotropy.

%%%%%%%%%%%%%%%%%%%%%%%%%%%%%%%%%%%%%%%%%%%%%%
\bibliographystyle{utphys}
\bibliography{main}
\end{document}